\documentclass[acmsmall]{acmart}

\renewcommand\footnotetextcopyrightpermission[1]{} 
\usepackage{lineno}

\usepackage{multirow}

\usepackage{amssymb}
\AtBeginDocument{%
  \providecommand\BibTeX{{%
    \normalfont B\kern-0.5em{\scshape i\kern-0.25em b}\kern-0.8em\TeX}}}

\begin{document}

\title{Motifs-based Recommender System via Hypergraph Convolution and Contrastive Learning}
\renewcommand{\shorttitle}{Motifs-based Recommender System via Hypergraph Convolution and Contrastive Learning.}


\author{Yundong Sun}
\orcid{0000-0001-5567-4553}
\email{hitffmy@163.com}
\author{Dongjie Zhu}
\authornote{Dongjie Zhu is the corresponding author.}

\email{zhudongjie@hit.edu.cn}

\author{Haiwen Du}
\email{duhaiwen@126.com}
\author{Zhaoshuo Tian}
\email{tianzhaoshuo@126.com}
\affiliation{
  \institution{Harbin Institute of Technology}
  \streetaddress{No.92,Xidazhi Street,Nangang District}
  \city{Harbin}
  \state{Heilongjiang}
  \country{China}
  \postcode{150001}
}

\renewcommand{\shortauthors}{Yundong Sun and Dongjie Zhu, et al.}

\begin{abstract}
Recently, the strategy of leveraging various motifs to model social semantic information and using self-supervised learning tasks to boost recommendation performance has been proven to be very promising. In this paradigm, each channel describes a common motif (e.g., a triangular social motif) via hypergraph convolution. Richer motifs can be encoded through multiple channels, and self-supervised learning can leverage this multichannel information to build self-supervised tasks (such as contrastive learning tasks), which can greatly improve the recommendation performance in scenarios without enough data labels. However, accurately determining the relationships between different channels and fully utilizing them, while maintaining the uniqueness of each channel, is a problem that has not been well studied or resolved in this field. This paper explores and verifies the disadvantages of directly constructing contrastive learning tasks on different channels with practical experiments and proposes a scheme of interactive modeling and matching representation across different channels. This is the first such attempt in the field of recommender systems, and we believe that this paper will inspire future self-supervised learning research based on multichannel information. To solve this problem, we propose a cross-motif matching representation model based on attentive interaction, which can efficiently model the relationships between cross-motif information. Based on this, we also propose a hierarchical self-supervised learning model that realizes self-supervised learning within and between channels, respectively, which improves the ability of self-supervised learning tasks to autonomously mine different levels of potential information. We have conducted abundant experiments, and various metrics on multiple public datasets show that the method proposed in this paper significantly outperforms the state-of-the-art methods, regardless of the general or cold-start scenario. In the model variant analysis experiment, the benefits of the cross-motif matching representation model and the hierarchical self-supervised model proposed in this paper are also fully verified.
\end{abstract}



\maketitle

\section{Introduction}
With the development of network technology, the popularization of mobile smart devices, and especially the emergence of short video platforms, such as Tiktok\footnote{https://www.tiktok.com/}and Kwai\footnote{https://www.kwai.com/}, information has shown explosive growth. Enormous amounts of information are stored on networks, but it is increasingly difficult for us to obtain valuable information for ourselves, which poses greater challenges to recommender systems\cite{wei2019mmgcn}. At the same time, the interactions between online users have become more frequent and complex, and the integration of rich social relationships into recommender systems to alleviate the problem of data sparsity has become a hot topic. Motif, as the specific local structure involving multiple nodes, is first introduced in \cite{milo2002network}. It has been widely used to describe complex structures or interactive information with specific patterns in a wide range of networks. Leveraging different motifs to model social semantic information and using self-supervised learning tasks to boost recommendation performance has been proven to be a very promising approach\cite{yu2021self}. In this paradigm, a common motif (e.g., a triangular social motif) can be encoded by each channel (a network branch, different branches capture different aspects of information) in the network via hypergraph convolution\cite{ji2020dual}, hence, multiple channels can encode richer higher-order relationships, and self-supervised learning can leverage this multichannel information to build self-supervised learning tasks, such as contrastive learning tasks that are encoded based on different multichannel information, which can greatly improve the recommendation performance in scenarios without enough data labels\cite{yu2021self,wu2021self,wang2021self}. Intuitively, contrastive learning can be considered as learning by comparing positive pairs of "similar" inputs and negative pairs of "dissimilar" inputs.

In recent years, graph neural networks (GNNs) have achieved unprecedented success in various tasks, such as node classification, link prediction, and graph classification, due to their strong ability to extract graph data structures, node attributes, and relationships between nodes\cite{zhang2020deep,zhu2020huna,xu2019cross}. At the same time, the application of GNNs in recommender systems has become a hot topic. Some studies construct bipartite graphs from interactive information between users and items and utilize GNNs to identify multihop relationships between users and users or items and items \cite{qu2019end,jin2020efficient}, thereby overcoming the limitations of the traditional collaborative filtering paradigms in capturing higher-order relationships. In addition, various methods add knowledge graphs\cite{cao2019unifying,wang2019kgat} and social information\cite{zhou2021group} as supplementary information to the interaction graphs, which further improves the performance of the recommender systems.

More information means more complex interactions. The interactions can be different social motifs between users\cite{feng2019hypergraph,yu2021socially} or purchase motifs between users and items\cite{yu2021self}. These complex and high-order interactions can provide richer information and help improve recommendation performance. Although stacked multilayer GNNs can capture the high-order relationships of multiple hops between nodes, they cannot capture the high-order relationships with specific motifs. The emergence of hypergraph neural networks (Hyper-GNNs) solved this problem. A hypergraph can connect nodes (more than two) with a specific relationship pattern by one edge. This natural structure enables the mining of specific motifs between nodes\cite{feng2019hypergraph,ji2020dual}.

Although the introduction of Hyper-GNNs and various interactive information with specific motifs has greatly improved the performance of recommendation models, this performance improvement requires an enormous amount of interactive data. For new users or items in a cold-start scenario, the recommendation performance is poor due to the lack of interactive data. At the same time, simply aggregating interaction information with specific motifs through different channels will also result in the loss of data characteristics from these channels. Self-supervised learning (SSL) can realize label-free learning with the help of SSL tasks, and it can also dig out potentially valuable information autonomously, thereby alleviating the cold-start problem. More importantly, it can fully mine and utilize the characteristics of information from different channels to improve the recommendation performance. In the fields of computer vision (CV) and natural language processing (NLP), various SSL methods\cite{he2020momentum,chen2020simple,devlin2018bert} have been used to overcome data label deficiencies and have achieved great success. However, due to the continuity of node attributes in interaction graphs and the complex relationships between nodes, it is difficult to directly apply data augmentation methods in the fields of CV and NLP to the recommendation\cite{wu2021self}.

Researchers have also conducted many studies on SSL for recommender systems. The available methods focus mainly on generating local substructures by randomly removing edges and sampling nodes\cite{wu2021self,you2020graph} to obtain training samples for contrastive learning, which is beneficial for capturing the local graph structures. However, there is richer interactive information in recommender systems, such as various social motifs between users and the purchase motifs between items. It is difficult to effectively mine the semantic information of multiple motifs by simply using random methods for data augmentation. Different channels can provide not only richer semantic information, which is beneficial for improving performance, but also new possibilities for SSL. Some of the latest methods\cite{yu2021self} consider the importance of semantic information from different motifs and use different channels to capture this information, and construct SSL tasks within each channel. However, these methods cannot fully utilize the relationships between different channels to construct SSL tasks. Simply constructing contrastive learning tasks for data from different channels by some methods (e.g., maximum mutual information) renders the data of each channel extremely homogeneous and causes them to lose their uniqueness (we also refer to this as homogeneity, and we prove this in exploratory experiments, see Preliminaries 3.3). Therefore, we need a new method for efficiently modeling the relationships between the interactive information of different channels to obtain cross-motif information that can be used to establish SSL tasks, so that recommendation performance can be boosted. Such a method will reduce the dependence on data labels and data volume and further improve the performance in cold-start scenarios.

Based on the above analysis, we propose a new framework named Motifs-Res (short for Motifs-based recommender system via hypergraph convolution and contrastive learning). First, inspired by MHCN\cite{yu2021self}, the data of three motifs in the interaction graph are extracted, namely, a social motif, purchase motif, and joint motif, and Hyper-GNN is leveraged to encode information of each motif. On this basis, we innovatively propose a cross-motif matching representation model that constructs first-order ego-networks of the nodes under each motif and realizes the information transition between different motifs based on the proposed attentive-matching representation learning model. More importantly, to solve the data homogeneity problem of each motif (Preliminaries 3.3), a hierarchical SSL model that is based on cross-motif matching representation is proposed, and a contrastive learning task is established for matching representations under different motifs.

The contributions of this paper can be summarized as follows:

\begin{itemize}
  \item [1.] We explore and experimentally verify the disadvantages of simply constructing contrastive learning tasks for data from different motifs and propose a scheme of interactive modeling and matching representation across different motifs. To the best of our knowledge, this is the first such attempt in the field of recommender systems, which can help facilitate subsequent self-supervised learning research based on multichannel information.   
  \item [2.] We propose a new framework named Motifs-Res (short for Motifs-based recommender system via hypergraph convolution and contrastive learning). Specifically, we propose a cross-motif matching representation model that is based on attentive interaction, which realizes efficient modeling of the relationships between cross-motif information. Based on this, we also propose a hierarchical self-supervised model, which realizes self-supervised learning within and between motifs, respectively, and improves the ability of self-supervised tasks to autonomously mine different levels of potential information.
  \item [3.] We conduct abundant experiments on various real datasets, and the results show that the proposed Motifs-Res method outperforms the state-of-the-art methods by a large margin in both general and cold-start scenarios. In a model variant analysis experiment, the benefits of the cross-motif matching representation model and the hierarchical self-supervised model that are proposed in this paper are also fully verified.
\end{itemize}

The remainder of this paper is organized as follows: Chapter 2 will introduce and analyze the research work that is related to this paper, including work on hypergraph neural networks, self-supervised learning, and self-supervised graph learning for recommendation. Chapter 3 will introduce definitions and basic technologies that are utilized in this paper, such as hypergraphs and hypergraph neural networks, and analyze our exploration experiments on self-supervised contrastive learning on multi-motif information. We will describe and analyze our method in detail in Chapter 4. Chapter 5 will present the experiments. We will evaluate the method that is proposed in this paper via performance comparison experiments, variant experiments, and stability experiments. Chapter 6 will present the conclusions of this paper and discuss future work.

\section{Related Work}

In this section, we will review and analyze the research works that are related to this paper.

\subsection{Hypergraphs and Hypergraph Neural Networks}

In recent years, the mining and analysis of graph data have become a research hotspot in the field of data mining and artificial intelligence. Graphs can well represent node attributes and topological structures, which can better model the affinities between nodes\cite{sun2022mhnf}. In particular, the emergence of GNNs has accelerated the industrialization process based on graph data research, such as knowledge question answering and dialogue systems\cite{wang2017knowledge,ji2021a}, recommender systems\cite{qu2019end,jin2020efficient,cao2019unifying,wang2019kgat,zhou2021group}, and intelligent search\cite{choudhary2021self,liu2020web}.

However, with the generation of more and more interactive data, a relationship may not only be limited to two nodes, that is, node pairs cannot represent some more complex interactive relationships, such as the stable triangular relationship in social networks\cite{yu2021socially}, the joint purchase relationship in the e-commerce networks\cite{yu2021self}, etc. The hypergraph can break this limitation of the traditional graph, its edge can connect any number of nodes, and is no longer limited to node pairs. Recently, some scholars have also begun to explore modeling methods on hypergraphs. Feng et al.\cite{feng2019hypergraph} propose a hypergraph neural network(Hyper-GNN) representation framework, called HGNN, which takes full advantage of the fusion ability of multi-channel data by combining adjacent matrices of hypergraphs. Compared with the traditional GNNs, it can better model the multi-channel interactive data of the social network. 

Some Hyper-GNN-based methods for recommender systems are proposed recently. Wang et al.\cite{wang2020next} pointed out that users may interact with different numbers of items in recommendation scenarios, but the traditional graph structure can only represent the pairwise relationship, which is not suitable for a real recommendation scenario. Therefore, they use sequential hypergraphs to model the user's behavior in different periods and use the Hyper-GNN to capture the interaction of multi-hop connections. In addition, through the combination with residual gating and fusion layer, the model can distill user preference more accurately, and the ability of sequence recommendation is significantly improved. Ji et al.\cite{ji2020dual} argue that whether traditional collaborative filtering recommendations that based on matrix factorization or graph-based collaborative filtering recommendations, there are deficiencies in the flexibility of modeling the high-order relationships between users and items. Based on this analysis, they propose a hypergraph-based collaborative filtering recommendation model, DHCF. The model adopts a dual-channel strategy and uses Jump Hyper-GNN (JHConv) to model users and items. The experimental results prove the value of high-order information for representation learning and the effectiveness of the proposed dual-channel hypergraph model. Valerio et al.\cite{la2022music} propose a new framework for song recommendation (HEMR) based on hypergraph embeddings. It utilizes a hypergraph model to seamlessly represent all possible and complex interactions between users and songs with relevant features. Meanwhile, it utilizes embedding techniques to infer user-song similarities through vector mapping. Results on the million-song dataset show that HEMR significantly outperforms other state-of-the-art techniques, especially in the case of cold-start, making the method an ideal solution for embedding in music streaming platforms. Our paper also uses Hyper-GNNs under multichannel to model the high-order interaction between users and items, but it is fundamentally different from the above methods. Our paper is devoted to modeling the interaction information from multi-channel with different motifs, not just interaction information within each channel. More importantly, the focus of this paper is to perform cross-motif matching representation learning to discover the potential association relationships between different interaction patterns, and then make full use of it for self-supervised learning.

\subsection{Self-Supervised Learning}

After many years of development, machine learning methods, especially deep learning methods, have achieved great success in image processing\cite{li2021survey}, natural language processing\cite{li2020survey} and other fields. The advantage of deep learning is to mine the valuable features of data from massive amounts of samples, but it requires a large amount of data and labels as input, which makes it difficult to apply deep learning models to some scenarios where data or label is scarce. The emergence of self-supervised learning (SSL) can well alleviate this problem. The core goal of SSL is to extract valuable information from the data itself or the association between data. When labels cannot be obtained, SSL can play the role of unsupervised learning; in scenarios where limited labels are available, SSL can play the role of pre-training or tuning\cite{xie2021self}. SSL was first applied to the field of image processing, such as image restoration, image denoising\cite{xie2020noise2same}, etc. SSL methods can be divided into the contrastive model and prediction model\cite{liu2021self}. The contrastive model usually uses the encoder to encode the data pairs, and then distinguishes the positive instances by calculating the encoding feature distance between the positive and negative pairs (such as the maximum mutual information). The input of the prediction model is generally a single instance. First, a certain method is used to construct a label (generally called as self-generated label), then the data is encoded and predicted, and finally, the loss between the predicted label and the generated label is calculated.

At present, with the continuous in-depth research on graph data, some SSL methods for graph data have been proposed. Shi et al.\cite{wang2021self} propose a contrastive learning framework, which encodes node representations from the network view and the meta-path view, respectively, which is used to establish a contrastive learning task. The framework improves the ability to extract both local and high-order structures of nodes. Qiu et al.\cite{qiu2020gcc} believe that the attributes of nodes are not transferable, but the structure can be transferred between different graphs. Therefore, they focus on studying how to mine the structural similarity of graphs through SSL tasks and transfer information between different graphs. The GCC framework they proposed is to sample the same nodes in the same ego-network to obtain positive subgraphs and use noise interference from other ego-networks to obtain multiple negative subgraphs. Different GNN-based encoders are used in these subgraphs, and the contrastive learning task is established to pre-training the encoders. The experimental results show that the GCC pre-training framework can greatly improve the performance of the initial model.

\subsection{Self-supervised Graph Learning for Recommendation}

At present, the graph-based recommendation model has become the most popular topic, which greatly boosts the performance of the recommendation. However, this performance improvement is guaranteed by enormous interactive information. On the one hand, with the higher real-time requirements of the recommender systems, the obstacle to obtaining the latest high-quality labels in time makes it difficult for the recommender systems to iteratively train in real-time\cite{ma2020temporal}. On the other hand, in large-scale application platforms, new users, or new items have made the cold start problem more serious\cite{qian2020attribute}. SSL can realize label-free learning with the help of self-supervised tasks, it can also improve the utilization of data, autonomously dig out potentially valuable information, and alleviate cold start problems in real-time recommendation scenarios to some extent.

The combination of SSL and recommender systems has become a new research hotspot in the past two years. Data augmentation is essential for SSL. Due to the continuity of node attributes and the complex relationships between nodes, data augmentation methods in the fields of CV and NLP are difficult to apply directly to recommendation applications\cite{wu2021self}. The SGL model\cite{wu2021self} uses edge dropout, node dropout, and random walks to generate structural variants of nodes from an original graph. This data augmentation strategy is conducive to capturing the structural pattern of the graph. However, the patterns of social relations between users in recommender systems can be diverse, as can those of the interactions between items. It is difficult to mine the semantic interaction information under various channels in recommender systems using a random strategy.

To explore the high-order interactions and semantic interaction patterns in recommender systems, some recent studies have begun to combine Hyper-GNNs with SSL. Xin et al.\cite{xia2020self} made the first attempt. First, they leveraged a hypergraph to construct the session association relationships between items; then, they constructed an association graph between sessions. The proposed two-channel Hyper-GNN was used to obtain feature representations of the two channels, based on which an SSL task was established. The experimental results show that the proposed SSL significantly improved the recommendation performance as an auxiliary task. Yu et al.\cite{yu2021self} proposed the MHCN model, which uses a hypergraph to encode the features of three channels: social motifs, joint motifs, and purchase motifs. Finally, in each channel, the proposed hierarchical mutual information maximization model is used as a self-supervised auxiliary task to optimize the recommendation model, which greatly improves the recommendation performance. The SEPT framework\cite{jin2020efficient} first adopts a similar strategy to extract the user triangle relationship from a social network and a user-item graph in the friend view and sharing view, respectively. It uses pseudo labels that are generated by other channels to perform SSL tasks on the predicted labels of this channel as an auxiliary task to optimize the recommendation model. Experimental results showed that the framework can optimize and improve previously proposed recommendation models (such as LightGCN\cite{he2020lightgcn}). A self-supervised hypergraph learning framework for group recommendation was proposed by Zhang et al.\cite{zhang2021double}. First, it develops a hierarchical hypergraph convolutional network based on user- and group-level hypergraphs to model correlations between them. Second, it designs a double-scale loss strategy for creating a self-supervised signal that can normalize user representations at different granularities to address sparsity issues. Experimental analysis on various datasets demonstrated the superior performance of the proposed model. The above models can mine high-order interaction relationships and semantic interaction patterns with the help of Hyper-GNNs and use these interaction patterns (channels) with SSL to optimize the recommendation task. However, these methods ignore the mining of the relationships between channels. Although SEPT uses the pseudo labels that are generated by other channels for SSL tasks, it does not fully mine the semantic matching relationships of different channels. This is also the essential difference between the method proposed in this paper and the previous methods.

\section{Preliminaries}
\subsection{Hypergraph Definition}

For a general graph, an edge can only connect two nodes, so it can only describe a pairwise relationship. But in real scenarios, many relationships are not limited to two nodes, such as the stable triangular relationship in social networks\cite{yu2021socially}. The hypergraph breaks this limitation. The hypergraph evolves an edge into a hyperedge, and a hyperedge can connect multiple nodes at the same time, which is also the core idea of the hypergraph\cite{feng2019hypergraph}. A hypergraph can be defined as $G = (V,E)$, where $V$ represents the set of nodes, and $E$ represents the set of hyperedges in the graph. The adjacency relationship between nodes can be represented as matrix $A \in {\{ 0,1\} ^{|V| \times |E|}}$ , where each element indicates whether hyperedge $e$ contains node $v$:
\begin{equation}
 a(v,e) = \left\{ {\begin{array}{*{20}{c}}
1&{if{\rm{ }}\ v \in e}\\
0&{if{\rm{ }}\ v \notin e}
\end{array}} \right.
\end{equation}

Like a general graph, the degree of a node in a hypergraph can be expressed by the number of hyperedges that contain the node, that is $d(v) = \sum\limits_{\forall e \in E} {a(v,e)}$. In addition, because a hyperedge is no longer limited to connecting two nodes, the degree of the hyperedge can be additionally defined as the number of nodes connected by it, that is $d(e) = \sum\limits_{\forall v \in V} {a(v,e)} $. Therefore, the diagonal matrices ${D_e} \in {\mathbb{R}^{|E| \times |E|}}$ and ${D_v} \in {\mathbb{R}^{|V| \times |V|}}$ formed by $d(e)$ and $d(v)$ represent the degree matrix of the hyperedges and nodes in hypergraph $G$, respectively.

\subsection{Extraction and Encoding of Information with Different Motifs}
\begin{figure}[t]
  \centering
  \includegraphics[width=\linewidth,trim=-10 20 -10 0]{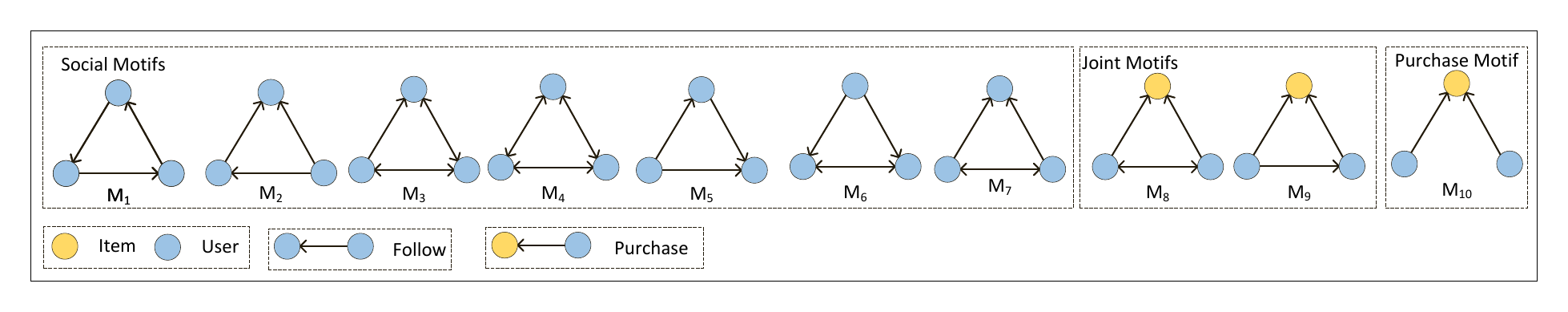}
  \caption{The triangle relationships extracted in this paper.}
  \vspace{-1.5em}
\end{figure}
Closely connected people are most likely to have the same interests. At present, with the popularity of various social platforms and the accumulation of massive data, it has become a trend to integrate users' social network information into the recommender systems\cite{yu2021socially,yu2021self}. The fusion of social network information and User-Item interactive information makes the relationship between user-user and user-item richer and no longer limited to pairwise relationships, which brings both advantages and challenges to relationship mining. How to extract and encode information with various patterns is not the focus of this paper. We follow the relevant content in MHCN\cite{yu2021self}.

First, we extract the relationship of the three patterns, namely Purchase Motif, Joint Motif, and Social Motif, as shown in Figure 1. A large number of studies have proven that triangles are extremely important in social relationships\cite{yu2021socially,benson2016higher}, among which Social Motif ($M_1$-$M_7$) covers the most important and stable triangle social relationships\footnote{We have also explored other triangle relations except for $M_1$-$M_7$ but found that adding these triangle relations does not improve the performance.}. Joint Motif ($M_8$ and $M_9$) means that two users have a social relationship and bought the same product. Purchase Motif ($M_{10}$) means that two users who do not have social relationships have purchased the same product. Although they have no social relationships, their demands or interests are similar. This is a very important potential relationship for recommendations.

\begin{table}[h]
\caption{The matrix computation of differnet motifs.}
\label{tab:motif}
\center
\vspace{-1.2em}
\begin{tabular}{c|l|l}
\toprule[0.4mm]
Motif & Matrix Computation & $A_{M_{i}}$             \\ \hline
$M_{1}$    & $C = (UU) \odot {U^ \top }$      & $C + {C^ \top }$ \\
$M_{2}$    & $C = (BU) \odot {U^ \top } + (UB) \odot {U^ \top } + (UU) \odot B$  & $C + {C^ \top }$ \\
$M_{3}$    & $C = (BB) \odot U + (BU) \odot B + (U \cdot B) \odot B$      & $C + {C^ \top }$ \\
$M_{4}$    & $C = (BB) \odot B$      & $C$ \\
$M_{5}$    & $C = (UU) \odot U + (U{U^ \top }) \odot U + ({U^ \top }U) \odot U$      & $C + {C^ \top }$ \\
$M_{6}$    & $C = (UB) \odot U + (B{U^ \top }) \odot {U^ \top } + ({U^ \top }U) \odot B$   &   $C$ \\
$M_{7}$    & $C = ({U^ \top }B) \odot {U^ \top } + (BU) \odot U + ({U^ \top }U) \odot B$      & $C $ \\
$M_{8}$    & $C = (R{R^ \top }) \odot B$      & $C $ \\
$M_{9}$    & $C = (R{R^ \top }) \odot U$      & $C + {C^ \top }$ \\
$M_{10}$   & $C = R{R^ \top }$      & $C $ \\ 
\bottomrule[0.4mm]
\end{tabular}
\vspace{-0.5em}
\end{table}

The corresponding adjacency matrix extraction method is shown in Table~\ref{tab:motif}. After obtaining the adjacency matrix ${A_{{M_K}}}$ under each motif, then the adjacency matrix of Social Motifs is ${A_s} = \sum\nolimits_{k = 1}^7 {{A_{{M_k}}}}$; the adjacency matrix of Joint Motifs is ${A_j} = {A_{{M_8}}} + {A_{{M_9}}}$; the adjacency matrix of Purchase Motifs is ${A_p} = {A_{{M_{10}}}} - {A_j}$.

After obtaining the adjacency matrix of the three motifs, the hypergraph convolutional neural network proposed in \cite{feng2019hypergraph}can be used to encode information of each motif:
\begin{equation}
 H_c^{(l + 1)} = \hat D_c^{ - 1}{A_c}H_c^{(l)}
\end{equation}
where ${D_c} \in {\mathbb{R}^{n \times n}}$ is the degree matrix of ${A_c}$.

\subsection{Empirical Explorations of Self-supervised Learning on Cross-Motif Information}

In this section, we explore leveraging different motifs to perform self-supervised learning. After obtaining representations of different motifs by different channels, an intuitive strategy is to use the contrastive loss function to make the representations between the channels of the positive samples closer and the representations of the negative samples more distant. Therefore, we incorporate Triplet-SSL loss\cite{dong2018triplet} and InfoNCE-SSL loss\cite{gutmann2010noise} into the $MHCN$\cite{yu2021self} to conduct contrastive learning on multichannel information directly, and format the $MHCN_{+Triplet}$ and $MHCN_{+InfoNCE}$ models. We keep the other hyperparameters unchanged to evaluate the performance of the model variants on different data. The exploration results are shown in Figure 2.

\begin{figure}[h]
  \centering
  \includegraphics[width=.5\linewidth,trim=0 20 0 0]{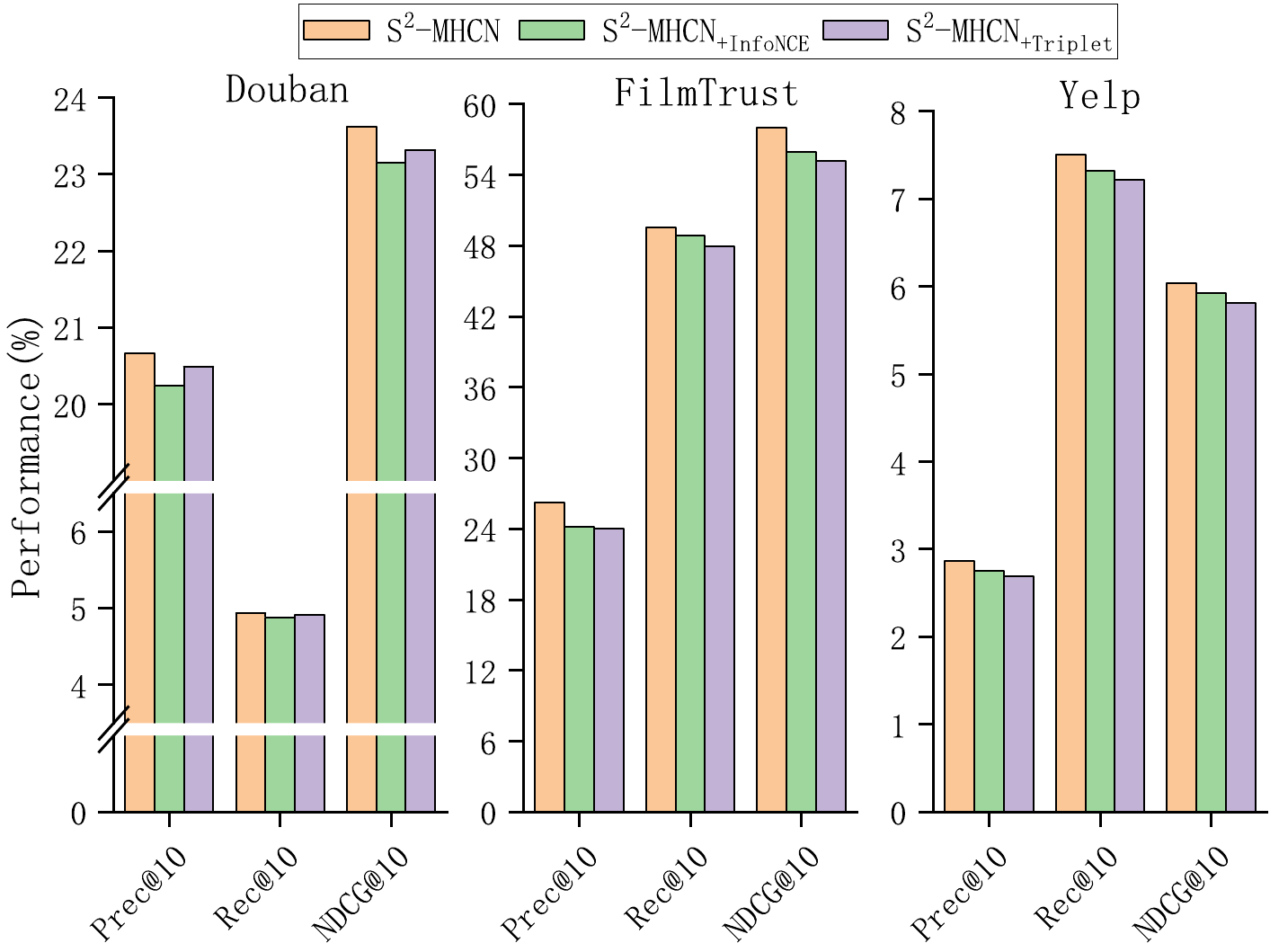}
  \caption{Performance exploration results of conducting contrastive learning on multichannel information directly.}
\end{figure}

Figure 2 shows that the performances of $MHCN_{+Triplet}$ and $MHCN_{+InfoNCE}$ on all metrics consistently decline compared with that of $MHCN$, which indicates that directly conducting contrastive learning on multichannel information is not a satisfactory scheme. This is in line with the original assumption of this paper: directly conducting contrastive learning on multichannel information will make the data of each channel extremely homogeneous and cause them to lose their uniqueness. This contrastive learning task not only fails to improve but also deteriorates the recommendation performance. Therefore, a new method is needed for efficiently modeling the relationships of different channels while maintaining their characteristics to maximize the advantages of SSL, which is the key problem that this paper solves.

\section{The Proposed Method}

Extensive recent research shows that leveraging different channels to model social semantic information and using self-supervised learning tasks to boost recommendation performance is a very promising approach. At the same time, the explorations and experiments that we conduct in Section 3.3 demonstrate the disadvantages of simply constructing contrastive learning tasks for the data from three motifs. Therefore, we need a scheme of interactive modeling and representation matching across motifs. Based on this, We propose a new framework named Motifs-Res (short for Motifs-based recommender system via hypergraph convolution and contrastive learning).

\begin{figure}[ht]
  \centering
  \includegraphics[width=\linewidth,trim=-10 30 -10 0]{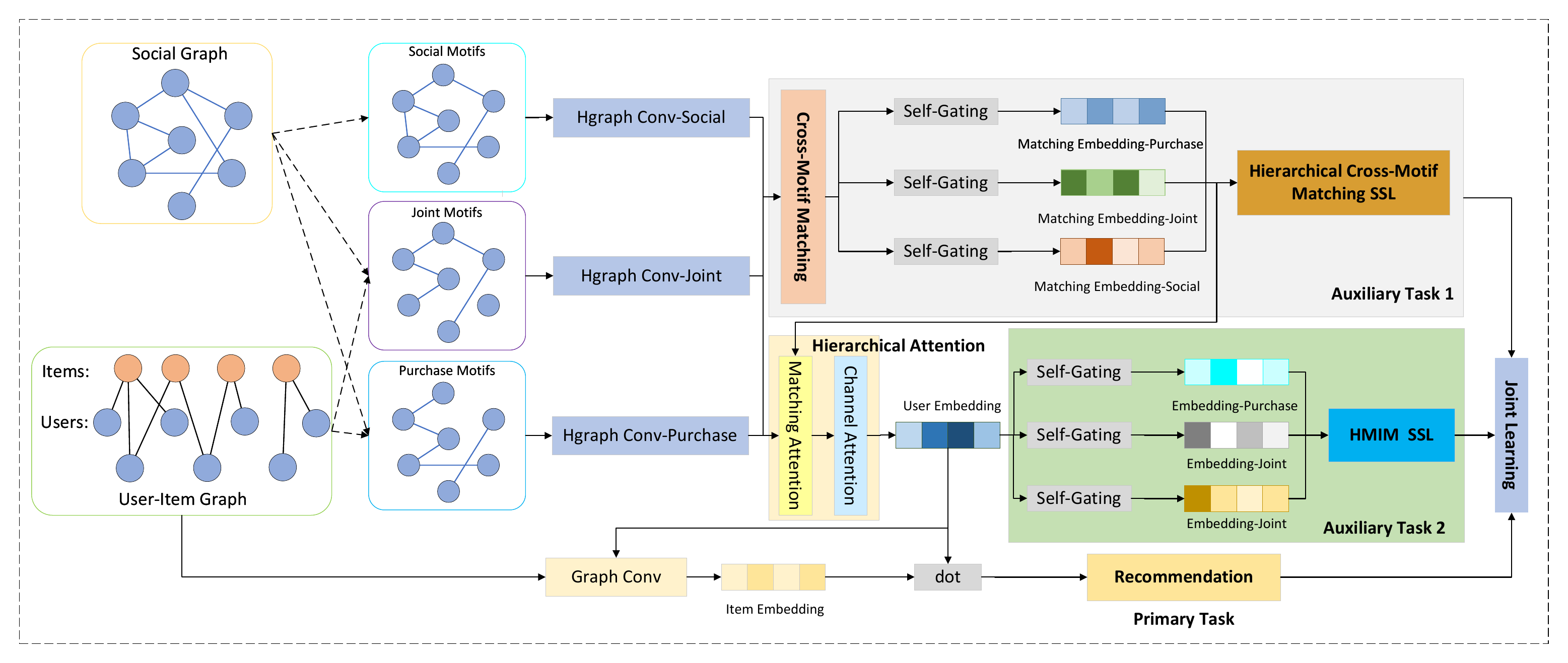}
  \caption{ Schematic diagram of the Motifs-Res framework (one layer) proposed in this paper.}
\end{figure}

This section will describe our proposed Motifs-Res framework in detail, and a diagram of its overall structure is shown in Figure 3. First, we construct three channels of information with various interaction patterns based on a social graph and a user-item graph, namely, a Purchase motif, Joint motif, and Social motif \cite{yu2021self,milo2002network} (Preliminaries 3.2). Then, the Hyper-GNN is used to encode the information of each channel to obtain the user representation under each motif. For each item, the GNN is implemented on the user-item graph to obtain its representation (Preliminaries 3.2). Finally, based on the user representations of each motif and the item representations, we conduct joint learning between the auxiliary task of the hierarchical SSL and the main task of recommendation. Cross-motif matching representation learning (Section 4.1), a cross-motif  contrastive learning model (Section 4.2), and user representation fusion based on hierarchical attention (Section 4.3) are the core methods of this paper.

\subsection{Cross-motif Matching Representation Learning}
\subsubsection{Framework of Cross-motif Matching Representation Learning.} We use the three channels as examples to illustrate and analyze the cross-motif matching representation framework that is proposed in this paper. A diagram of the overall architecture is shown in Figure 4 (a). First, for the data of any two motifs, we use the cross-motif representation learning model (the attentive-matching model) proposed in this paper to obtain the transitional matching representation from motif 2 to motif 1 or from motif 1 to motif 2. Then, the transitional matching representations from the other two motifs to the current motif are summed to obtain the cross-motif transitional representation of the current motif. For example, the cross-motif matching representation of the social motif can be expressed as follows:
\begin{equation}
H_s^m = Attm(H_s^n,H_j^n) + Attm(H_s^n,H_p^n)
\end{equation}
where $H_s^n,H_j^n,H_p^n$ denote the user representations under the three motifs that are obtained by Equation (2) and $Attm(\cdot)$ is the attentive-matching model that we propose in Section 4.1.2. We extend it to a more general case (with $L$ motifs), and the matching representation of each motif is calculated as follows:
\begin{equation}
H_i^m = \sum\limits_{0 < l \le L,l \ne i} {Attm(H_i^n,H_l^n)} ,i = 1,2,...,L
\end{equation}
where $H_i^n$ is the user representation under motif $i$, and $H_i^m$ is the cross-motif matching representation of the user under motif $i$.

\begin{figure}[t]
  \centering
  \includegraphics[width=.6\linewidth,trim=0 50 0 20]{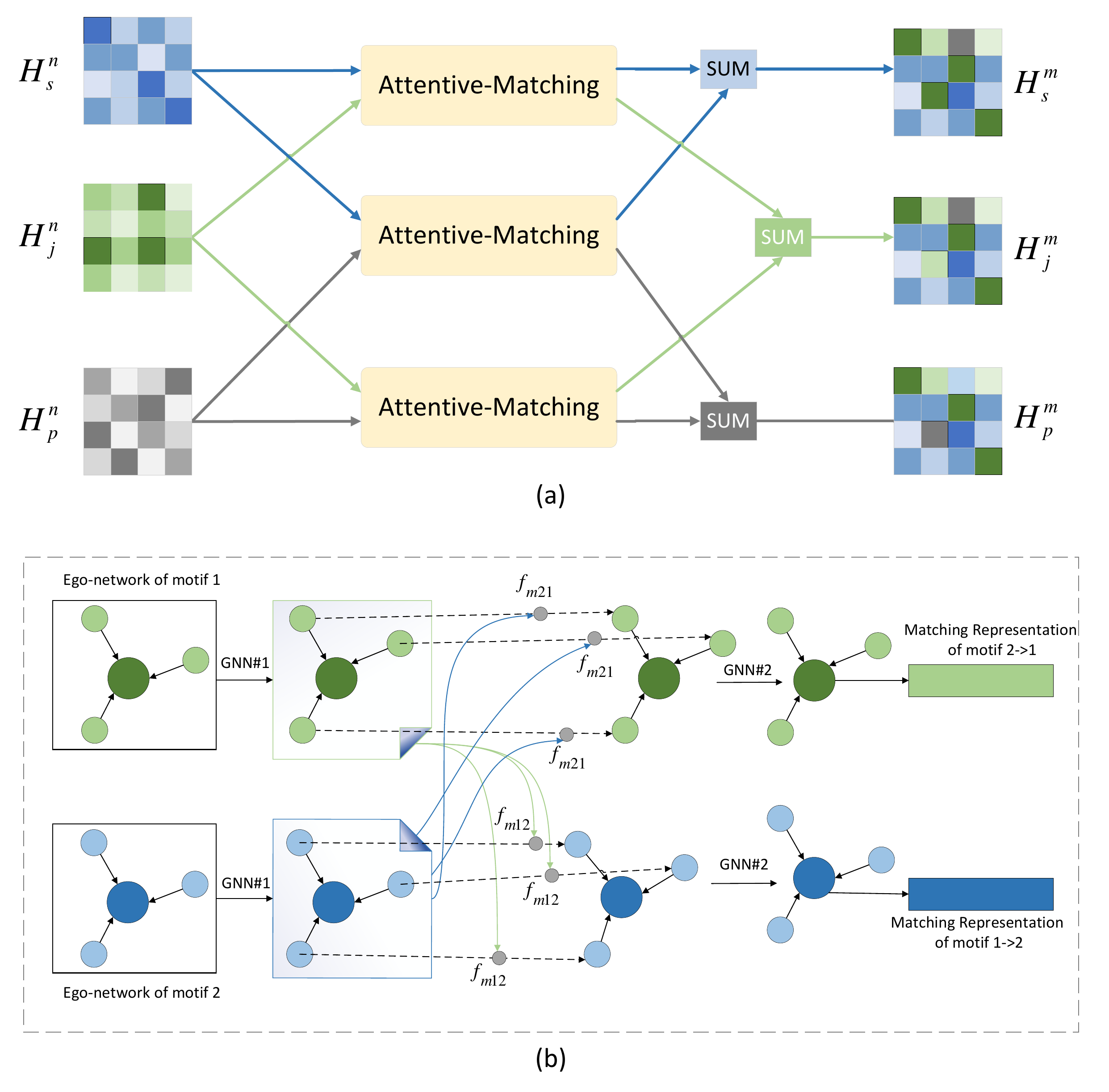}
  \caption{ (a) The framework of cross-motif matching representation learning and (b) a detailed illustration of the attentive-matching model.}
  \vspace{-1.3em}
 \end{figure}

\subsubsection{Attentive-Matching Model.} To obtain a transitional matching representation between two motifs, this paper proposes an attentive-matching cross-motif representation learning model. First, we use $GCN\#1$\cite{kipf2016semi} to gather information on the ego-network under each motif to obtain the ego-network representations $g_i^1$ an $g_j^2$ of the node $i$ and $j$.

On this basis, the nodes of the ego network under other motifs are used to perform cross-motif transitional representation modeling for the nodes of the current motif. Specifically, we first calculate the cosine similarity between $g_i^1$ and $g_j^2$, as expressed in Equation (5).
\begin{equation}
{\alpha _{i,j}} = \cos ine(g_i^1,g_j^2),j \in 1,...,|{G_2}|
\end{equation}

Then, we perform a weighted summation on the representations of all nodes in ${G_2}$ to obtain the transitional representation of the nodes in ${G_1}$, as expressed in Equation (6).
\begin{equation}
g_i^{ - 1} = \frac{{\sum\nolimits_{j = 1}^{|{G_2}|} {{\alpha _{i,j}} * g_j^2} }}{{\sum\nolimits_{j = 1}^{|{G_2}|} {{\alpha _{i,j}}} }}
\end{equation}

Finally, we utilize representation $g_i^1$ and transitional representation $g_i^{ - 1}$ to obtain the transitional matching representation of node $i$ from motif 2 to motif 1, as expressed in Equation (7).
\begin{equation}
h_i^m = {f_m}(g_i^1,g_i^{ - 1})
\end{equation}
where ${f_m}$ is a multiview cosine matching function that compares two vectors of ${v_1}$ and ${v_2}$:
\begin{equation}
{h^m} = {f_m}({v_1},{v_2},W)
\end{equation}
where $W \in {\mathbb{R}^{l * d}}$ is a trainable parameter, $l$ is a multiview parameter, and $W_k$ denotes one row of $W$, which represents a single view. $d$ is the dimension of the input vectors, and $m = [{m_1},{m_2},...,{m_k},...,{m_l}]$ is an $l$-dimensional matching vector. In special cases, if $l=1$ , it is the common vector cosine similarity of the input vectors:
\begin{equation}
{m_1} = cosine({W_1} * {v_1},{W_1} * {v_2})
\end{equation}

To better capture the local structural information of the nodes, $GCN\#2$\cite{kipf2016semi} is used again to gather the ego-network information to obtain the final cross-motif matching representation under each motif. We apply the same process to the nodes in ${G_2}$.  

\subsection{Hierarchical Self-supervised Learning based on Cross-motif Matching Representation.}
In Section 4.1, we obtained the transitional matching representation between different motifs. The matching representation contains the association relationships between different motifs, which become the basis for applying contrastive learning among motifs. At the same time, similar to general information, the matching representations under each motif still maintain different distributions, which have unique characteristics. To preserve the uniqueness of the respective motifs and make full use of the correlations between the matching representation from different motifs, this paper proposes a hierarchical self-supervised learning model based on cross-motif matching representation. A diagram of the model is shown in Figure 5.

 \begin{figure}[t]
  \centering
  \includegraphics[width=\linewidth,trim=0 30 0 20]{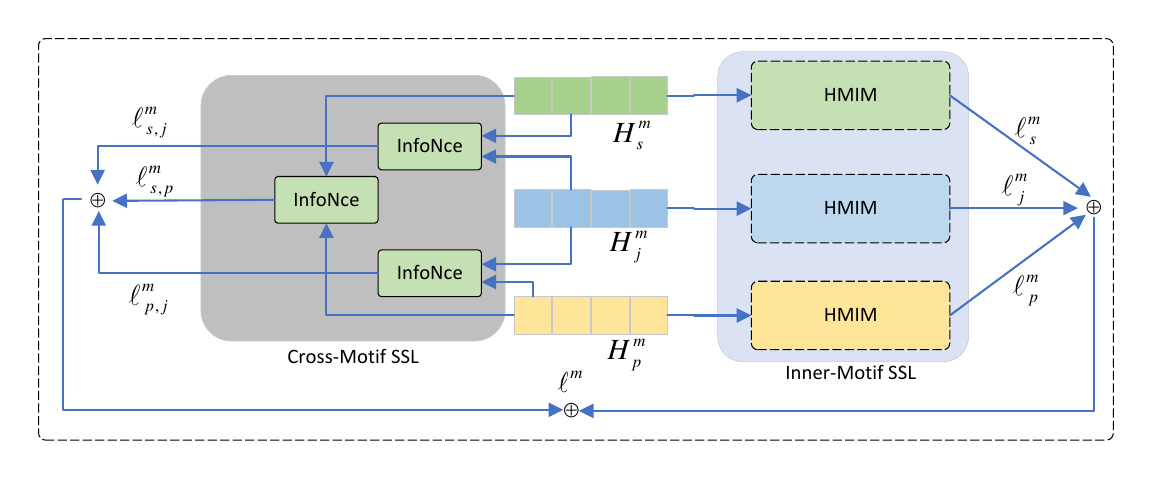}
  \caption{Schematic diagram of hierarchical self-supervised learning based on cross-motif matching representation.}
  \vspace{-1.5em}
 \end{figure}
First, we build a contrastive learning task on representations of multiple motifs. The commonly used objective functions, such as the maximum mutual information model\cite{velickovic2019deep} and the InfoNCE model\cite{haresamudram2021contrastive}, can correctly capture the similarity representation of instances between different motifs.

Specifically, in each calculation process, one positive sample (${k_ +}$) and $K$ negative samples (${k_0},...,{k_K}$) from the queue are selected for calculation, and InfoNCE\cite{haresamudram2021contrastive} can be used to perform the following calculation:

\begin{equation}
\ell  =  - \log \frac{{\exp (h_q^ \top h_{_{{k_ + }}}^m/\tau )}}{{\sum\nolimits_{i = 0}^K {\exp (h_q^ \top h_{{k_i}}^m/\tau )} }}
\end{equation}
where ${h_q}$ is the representation of the query node, which is the matching representation of node $q$ under a motif that was obtained in Section 4.1, $h_{{k_i}}^m$ is the matching representation of the positive sample, $h_{{k_i}}^m$ is the matching representation of the $i^{th}$ negative sample, and $\tau$ is the temperature hyperparameter.

In practical implementation, to improve the computational efficiency and save GPU memory, we don't set the cache queue of negative samples as in the MoCo strategy\cite{he2020momentum} but instead perform operations on instances within the same minibatch. In terms of matrix operations, this can be expressed as follows:

\begin{equation}
\ell _{i,j}^m =  - \log \frac{{\exp [(\sum\limits_{colum} {(H_i^m*H_j^m))} /\tau ]}}{{\exp [(\sum\limits_{colum} {(H_i^m \times H{{_j^m}^ \top }))} /\tau ]}}
\end{equation}
where $H_i^m \in {\mathbb{R}^{n \times d}}$ and $H_j^m \in {\mathbb{R}^{n \times d}}$ represent the matching representations of the nodes in the current minibatch under motif $i$ and $j$, respectively. $n$ is the size of the minibatch, $*$ is the Hadamard product, $\times $ is matrix multiplication, $\sum\limits_{colum}A$ represents columnwise summation of the two-dimensional matrix $A$, and $\tau$ is the temperature hyperparameter.

Therefore, the final self-supervised contrastive loss function based on cross-motif matching under the three motifs is:
\begin{equation}
{\ell _{s11}} = \ell _{s,j}^m + \ell _{s,p}^m + \ell _{p,j}^m
\end{equation}

To preserve the characteristics of the matching representation under different motifs, SSL based on maximum mutual information is performed in each motif. According to Yu et al.\cite{yu2021self}, the maximum mutual information model that is used in DGI\cite{velickovic2019deep} is established directly between the target node representation and the motif graph representation. This is a relatively coarse-grained strategy, which cannot guarantee that the encoder distills sufficient information from the inputs. For a large-scale graph, there is a more significant difference between the node representation and the entire graph, and the maximum mutual information will lose its effect and even play a negative role. Therefore, this paper adopts the HMIN model, which was proposed by Yu et al. \cite{yu2021self}, to establish a two-level maximum mutual information model in each motif. The first level is the maximum mutual information from a node to its corresponding ego-network, and the second level is from the ego-network to the corresponding graph. Therefore, we define the self-supervised loss function in a single channel as:
\begin{small}
\begin{equation}
{\ell _{s12}} =  - \sum\limits_{i \in \{ s,j,p\} } {\{ \sum\limits_{u \in U'} {\log \sigma [(h_{u,i}^m * z_{u,i}^m)}  - (\tilde h_{u,i}^m * z_{u,i}^m)] + \sum\limits_{u \in U'} {\log \sigma [(z_{u,i}^m * z_i^m)}  - (\tilde z_{u,i}^m * z_i^m)]\} } 
\end{equation}
\end{small}
where $U'\subset U$ is the user set of the current minibatch, $h_{u,i}^m \in {\mathbb{R}^d}$ is the matching representation of user $u$ motif $i$, $z_{u,i}^m = \sum\limits_{v \in N_u^i \cup \{ u\} } {e_{uv}^i * h_v^{{m_i}}}$ is the matching representation of the ego-network of user $u$ under motif $i$, and $e_{uv}^i$ denotes the weight of the edge between $u$ and $v$ under motif $i$. $\tilde h_{u,i}^m$ is the negative sample of $h_{u,i}^m$, which is obtained by the rowwise or columnwise shuffling of $h_{u,i}^m$, as is $\tilde z_{u,i}^m$.   $z_i^m = AveragePooling(H_i^m)$ is the average matching representation of all nodes in the graph under motif $i$. We sum the SSL losses of the inner-motif and the cross-motif to obtain the comprehensive SSL loss of the matching representation. We regard it as the loss of auxiliary task 1:
\begin{equation}
{\ell _1} = {\ell ^m} = {\ell _{s11}} + {\ell _{s12}}
\end{equation}
 
\subsection{Hierarchical Attention for Comprehensive User Representation} 

In this section, we need to fuse two levels of user information, namely, the common representation and matching representation of each user within one motif, and the information of different motifs. An intuitive strategy is to simply perform summation, averaging, or maximization, but the valuable information for the target task cannot be distilled for each user during the training process. Therefore, this paper proposes a hierarchical attention information fusion model in this section, which learns to extract the most valuable information from multiple levels.

First, we need to fuse the common representation and matching representation of users under each motif. For each user, we learn the attention coefficients $(\alpha _s^m,\alpha _s^n),(\alpha _j^m,\alpha _j^n),(\alpha _p^m,\alpha _p^n)$, which represent the weights of the user's common representation and matching representation under the three motifs. The attention function in this section can be defined as follows:
\begin{equation}
\alpha _c^q = {f_{att1}}(h_c^q) = \frac{{\exp ({a^ \top } \cdot {W_{att1}}h_c^q)}}{{\sum\nolimits_{q \in \{ m,n\} } {\exp ({a^ \top } \cdot {W_{att1}}h_c^q)} }}
\end{equation} 
where $a \in {\mathbb{R}^d}$ and ${W_{att1}} \in {\mathbb{R}^{d \times d}}$ are learnable parameters. Finally, the user representation under each motif is expressed as
\begin{equation}
{h_c} = \sum\nolimits_{q \in \{ m,n\} } {\alpha _c^qh_c^q}
\end{equation}  

Second, after obtaining the user representation under each motif, the user representations of different motifs need to be fused. Similar to user information fusion within one motif, we adopt a multichannel fusion technology based on the attention mechanism to obtain the comprehensive user representation of the three motifs: 
\begin{equation}
h = \sum\nolimits_{c \in \{ s,j,p\} } {{\alpha _c}{h_c}}
\end{equation}
where ${\alpha _c}$ is the attention coefficient of the corresponding motif and its learning method is similar to Equation (15).  
 
\subsection{Overall Optimization} 
 
After obtaining the user's comprehensive representation (Section 4.3), we use the interaction matrix $R$ between the users and the items to obtain the final item representation: 
\begin{equation}
{Q^{(l + 1)}} = D_i^{ - 1}{R^ \top }{H^{(l)}}
\end{equation}
where ${D_i} \in {\mathbb{R}^{n \times n}}$ is the degree matrix of $R$. 

Except for the hierarchical self-supervised learning tasks (auxiliary task 1), which are based on cross-motif matching representation, we refer to the MHCN\cite{yu2021self}, and leverage the HMIM model for self-supervised learning, as auxiliary task 2:
\begin{equation}
{\ell _2} =  - \sum\limits_{i \in \{ s,j,p\} } {\{ \sum\limits_{u \in U} {\log \sigma [(h_{u,i}^n * z_{u,i}^n)}  - (\tilde h_{u,i}^n * z_{u,i}^n)] + \sum\limits_{u \in U} {\log \sigma [(z_{u,i}^n * z_{\rm{i}}^n)}  - (\tilde z_{u,i}^n * z_{\rm{i}}^n)]\} }
\end{equation}
where $U$ is the user set of the current minibatch, $h_{u,i}^n \in {\mathbb{R}^d}$ is the common representation of user $u$ under motif $i$, $z_{u,i}^n = \sum\limits_{v \in N_u^i \cup \{ u\} } {e_{uv}^i * h_{v,i}^n}$ is the common representation of the ego-network of user $u$ under motif $i$, and $e_{uv}^i$ indicates the weight of the edge between $u$ and $v$ under motif $i$. $\tilde h_{u,i}^{\rm{n}}$ is the negative sample of $h_{u,i}^n$, which is obtained by rowwise or columnwise shuffling of $h_{u,i}^n$, as is $\tilde z_{u,i}^m$. $z_i^n = AveragePooling(H_i^n)$ is the average of the common representations of all nodes in the graph under motif $i$.
\begin{table}[t]
\label{tab:ALGORITHM}
\center
\begin{tabular}{l}
\toprule[0.4mm]
\textbf{ALGORITHM 1:}\hspace*{0.4in} \textbf{The Motifs-Res algorithm.}  \\ \hline
\textbf{Input:}\hspace*{0.1in}  User-Item graph adjacency matrix $R$ and social graph adjacency matrix $A$. \\ 
\textbf{Output:} Recommendation prediction result $\hat R$. \\ \hline
\hspace*{0.1in} 1. \hspace*{0.1in} All parameters are initialized. \\
\hspace*{0.1in} 2. \hspace*{0.1in} The relationship ${A_s},{A_j},{A_p}$ between users are obtained according to Table 1.  \\

\hspace*{0.1in} 3. \hspace*{0.1in} \textbf{for} i=1,2,3,...,$epoch$ \textbf{do}
 \\
\hspace*{0.1in} 4. \hspace*{0.1in}\hspace*{0.1in} The common representation $H_s^n,H_j^n,H_p^n$ are obtained according to Equation (2).  \\

\hspace*{0.1in} 5. \hspace*{0.1in}\hspace*{0.1in}  The matching representation $H_s^m,H_j^m,H_p^m$ are obtained according to Equation (4).\\

\hspace*{0.1in} 6. \hspace*{0.1in}\hspace*{0.1in}   The loss ${\ell _{s11}}$ of cross-motif contrastive learning based on matching \\ \hspace*{0.48in} representation is obtained according to Equation (13). \\

\hspace*{0.1in} 7. \hspace*{0.1in}\hspace*{0.1in}   The loss ${\ell _{s12}}$ of inner-motif contrastive learning based on matching \\ \hspace*{0.48in} representation is obtained according to Equation (14). \\

\hspace*{0.1in} 8. \hspace*{0.1in}\hspace*{0.1in}   The overall loss ${\ell _1} = {\ell ^m} = {\ell _{s11}} + {\ell _{s12}}$ of contrastive learning based on   matching \\ \hspace*{0.48in} representation  is obtained (Auxiliary task 1). \\
\hspace*{0.1in} 9. \hspace*{0.1in}\hspace*{0.1in}   The coefficients $(\alpha _s^m,\alpha _s^n),(\alpha _j^m,\alpha _j^n),(\alpha _p^m,\alpha _p^n)$ of user's normal and  matching \\ \hspace*{0.48in} representation under different motifs are obtained according to Equation (16). \\

\hspace*{0.1in}10. \hspace*{0.1in}\hspace*{0.1in}   The user representation ${h_s},{h_j},{h_p}$ under each motif are obtained by Equation (17).  \\

\hspace*{0.1in}11. \hspace*{0.1in}\hspace*{0.1in}  The user comprehensive representation $h$ over multi-motif is obtained \\ \hspace*{0.48in} according to Equation (18).   \\

\hspace*{0.1in}12. \hspace*{0.1in}\hspace*{0.1in}   The final item representation $Q$ is obtained according to Equation (19).  \\

\hspace*{0.1in}13. \hspace*{0.1in}\hspace*{0.1in}   The overall loss  ${\ell _2}$ of contrastive learning  is obtained based on the user common \\ \hspace*{0.48in} representation (Auxiliary task 2). \\

\hspace*{0.1in}14. \hspace*{0.1in}\hspace*{0.1in}   The recommendation prediction result $\hat R$ is obtained. \\

\hspace*{0.1in}15. \hspace*{0.1in}\hspace*{0.1in}   The loss of recommendation prediction ${\ell ^r}$ is obtained (Main task).   \\

\hspace*{0.1in}16. \hspace*{0.1in}\hspace*{0.1in}   The loss of joint learning $L$ is obtained.   \\

\hspace*{0.1in}17. \hspace*{0.1in}\hspace*{0.1in}  The Adam optimizer and backpropagation algorithm are used to optimize \\ \hspace*{0.48in} the model parameters.     \\
\hspace*{0.1in}18. \hspace*{0.1in} \textbf{end for} \\
\hspace*{0.1in}19. \hspace*{0.1in} \textbf{Return} $\hat R$.\\
  \bottomrule[0.4mm]
\end{tabular}
\end{table}

For the main task, the BPR loss function\cite{rendle2009bpr} is used to optimize the model:
\begin{equation}
{\ell _r} = \sum\limits_{i \in I(u)j \notin I(u)} { - \log \sigma ({{\hat r}_{u,i}} - {{\hat r}_{u,j}})}  + \lambda ||\theta ||_2^2
\end{equation}
where $I(u)$ is the set of items that are purchased by user $u$, $\theta $ is the model parameter, and $\lambda $ is the regularization coefficient. In each training process, a randomly selected positive sample item $i$ that is purchased by user $u$ and an undetected item $j$ form a triple for training optimization. ${\hat r_{u,i}}$ is the recommendation prediction between $u$ and $i$, which is calculated based on the representation of the user and item:
\begin{equation}
{\hat r_{u,i}} = h_u^ \top {q_i}
\end{equation}

In summary, auxiliary tasks 1 and 2 are jointly trained with the main task to optimize the model:
\begin{equation}
L{\rm{ = }}{\ell _{\rm{r}}} + {\beta _1}{\ell _1} + {\beta _2}{\ell _2}
\end{equation}
where ${\beta _1}$ and  ${\beta _2}$ are the coefficients of auxiliary task 1 and auxiliary task 2, respectively. Algorithm 1 describes the main steps that are applied to optimize the framework.

\section{Experiments}
In this section, we will conduct various experiments (including performance comparison experiments, ablation experiments, and parameter sensitivity experiments) on real datasets to answer the following questions:
\begin{itemize}
\item \textbf{RQ1:} How does the method proposed in this paper compare with the state-of-the-art recommendation methods?
\item \textbf{RQ2:} What are the benefits of the cross-motif matching representation and hierarchical SSL on cross-motif information proposed in this paper?
\item \textbf{RQ3:} How do the hyperparameter settings affect the model?
\end{itemize}

\subsection{Experimental Protocol}
\subsubsection{Datasets.}In the experiment, we select three public real datasets that are commonly used in recommender systems, namely, Douban\cite{zhao2016user}, FilmTrust\cite{golbeck2006filmtrust}, and Yelp\cite{yu2021self}, to evaluate various methodologies. The task of the experiment is to perform top-10 recommendations on the processed open-source data from\cite{yu2021self,rendle2009bpr}. Statistical information on the experimental data is presented in Table 2.
\vspace{-1em}
\begin{table}[h]
\caption{The statistical information of the experimental data.}
\label{tab:data}
\vspace{-1em}
	
\center

\begin{tabular}{c c c c c c}
\toprule[0.4mm]
Dataset 	& \#Users & \#Items  &  \#Ratings   & \#Relation   &  Density  \\ \hline
Douban	 	& 2,848  &  39,586   & 894,887   &  35,770 & 0.794\% \\ 
FilmTrust	 	& 1,508  &  2,071   & 35,500   &  1,854 &1.137\% \\ 
Yelp	 	& 19,539  &  21,266   & 450,884   &  864,157	 &0.11\% \\ 
  \bottomrule[0.3mm]
\end{tabular}
  \vspace{-1em}
\end{table}
\subsubsection{Baselines.}
To fully evaluate the performance of our proposed Motifs-Res framework, we compare it with the most advanced and commonly used baseline methods, which include MF-based methods, GNN-based methods, methods with SSL auxiliary tasks, and methods without SSL auxiliary tasks. Notably, SGL\cite{wu2021self} and $S^{2}$-MHCN\cite{yu2021self} are SOTA methods, and the following baseline methods are from the open-source library QRec\footnote{https://github.com/Coder-Yu/QRec}.

\begin{itemize}
\item BPR\cite{rendle2009bpr}. This is a classic recommendation method. It was first proposed as a general optimization method based on Bayesian personalized ranking. This paradigm has been used in subsequent research and is the basis for ranking recommendation algorithms.
\item SBPR\cite{zhao2014leveraging}. This is an optimized version of the BPR model\cite{rendle2009bpr}. It integrates social relationships into the BPR model and tends to assign higher rankings to items that have been liked by the friends of the target user. Compared with BPR, it achieves better performance in both general and cold-start scenarios.

\item NGCF\cite{wang2019neural}. This is a collaborative filtering recommendation algorithm based on GNN. It operates under the assumption that high-order relationships of the user and items contain rich collaborative filtering information. Its core strategy is to incorporate high-order relationships into the process of representation learning to better integrate collaborative filtering information into users and items.
\item DiffNet\cite{wu2019neural}. This is also a representative model of integrating social relationships into recommendations. However, it assumes that the social influence of the user is dynamic, and this influence causes the interests of involved users to constantly change during the process of diffusion. Based on this, a social influence communication model is designed to improve the performance of the recommendation model.
\item LightGCN\cite{he2020lightgcn}. This method was developed in a recent novel study. In this work, the author studied GNN-based collaborative filtering models such as NGCF and found that the two standard operations (model-feature conversion and nonlinearity operations) in the GNN are not necessary for collaborative filtering but increase the difficulty of training. Therefore, by simplifying the GNN model, they improved the recommendation performance of collaborative filtering while reducing the training difficulty and computational complexity.

\item SGL\cite{wu2021self}.  This is the latest method to use SSL to overcome the limitations of GNN-based recommendations under the supervised learning paradigm. It is believed that available GNN-based recommendation methods for learning representations of user-item graphs have limitations in solving long-tail recommendations and resisting noise interactions. Three strategies, namely, Node-Drop, Edge-Drop, and Random-Walk, are used to achieve data augmentation and obtain different graph structure views to construct SSL auxiliary tasks, which can effectively overcome the limitations of GNNs. However, the patterns of social relations between users in recommender systems can be diverse, as can those between items. It is difficult to mine the semantic interaction information under various channels in recommender systems by a random strategy. This is also a key problem to be solved in this paper.
\item $S^{2}$-MHCN\cite{yu2021self}. This is the latest social recommendation model with self-supervised learning. It combines social networks with user-item graphs, constructs three-motif information, and encodes each motif through a hypergraph neural network. On this basis, it performs self-supervised learning through its proposed HMIN model in each motif, which is regarded as an auxiliary task. However, it does not fully explore the relationships between different motifs, which is a significant difference from this paper.
\end{itemize}

\subsubsection{Metrics.} To comprehensively evaluate the models, we select three commonly used metrics in recommender systems: Precision@10, Recall@10, and NDCG@10. 

 \subsubsection{Experimental Settings.} For the baseline methods, we use grid search to determine the best parameter settings. The maximum iteration is searched in \{10,20,30,40,50,60,70,80,90,100\}, for other parameters, we use the best settings suggested in the original paper to maintain its optimal performance in real scenarios\footnote{To make the result more reasonable, we unify the parameters under all datasets instead of setting the corresponding optimal parameters for a specific dataset. For example, on the FilmTrust dataset, we don't increase the layer of Motifs-Res to improve the metric values, as shown in Figure 10.}. For the general settings of all methods involved in the experiment, the embedding (both user and item) dimension $d$ is set to 50, and the batch size is set to 2000. For our method, the regularization coefficient $\lambda $ is set to 0.03, the SSL loss coefficient ${\beta _1},{\beta _2}$ are searched in \{0.001,0.002,0.003,0.005,0.01,0.02,0.05,0.1\}, and finally, the optimal values are determined as ${\beta _1}{\rm{ = }}0.01$  and ${\beta _2}{\rm{ = }}0.001$. The temperature smoothing coefficient $\tau $ is searched in \{0.01,0.02,0.05,0.1,0.2,0.5,0.6,0.7,0.8\}, and the optimal value is finally determined as $\tau $ =0.5.  The maximum number of iterations is determined as 30, and the model layer is set to 2. To ensure the stability, comprehensiveness, and credibility of the results, the result of each experiment is the average result of five-fold cross-validation over 10 repeats. Our demo code and the used datasets are publicly available at \url{https://github.com/PHD-lanyu/Motifs-Res}.
 
\subsection{Performance Comparison(RQ1)}
 
In this section, we use performance comparison experiments to determine whether the Motifs-Res framework can outperform the most advanced recommendation methods. As analyzed previously, self-supervised learning can not only boost the recommendation performance in general scenarios by mining richer information but also alleviate the problem of data sparseness in cold-start scenarios with the help of the correlations between multi-motif information. Therefore, we evaluate the performances of the selected methods on complete datasets in a general scenario and on sparse data in a cold-start scenario. The sparse data in the cold start scenario only include users with fewer than 20 interactions. The experimental results are presented in Table 3 and Table 4, respectively. The best results are presented in bold; suboptimal results are underlined. From these results, we draw the following conclusions:

\begin{table}[]

\caption{ General recommendation performance result.}
\label{tab:result1}
\vspace{-1em}
\setlength\tabcolsep{2pt}

\begin{tabular}{cc|cccccccc|c}

\toprule[0.4mm]
Dataset                 & {Metric} & {BPR}     & {SBPR}    & {NGCF}     &{DiffNet}    & {SGL} &\footnotesize{LightGCN}   &\footnotesize{$S^{2}$-MHCN} & {\textbf{Motifs-Res}}   &{Improv.} \\
\midrule[0.3mm]

\multirow{3}{*}{Douban} &{P@10}   & {14.45\%} &{18.50\%} & \multicolumn{1}{c}{18.34\%} & \multicolumn{1}{c}{17.19\%} & \multicolumn{1}{c}{19.45\%} & \multicolumn{1}{c}{20.46\%}  & \multicolumn{1}{c} {\underline{20.76\%}} & {\textbf{21.18\%}} &{+2.02\%} \\

& {R@10}   &{2.98\%}  & \multicolumn{1}{c}{4.30\%}  & \multicolumn{1}{c}{4.30\%}  & \multicolumn{1}{c}{4.13\%}  & \multicolumn{1}{c}{5.00\%}   & \multicolumn{1}{c}{\textbf{5.24\%}}  & \multicolumn{1}{c}{4.82\%}  & {\underline{5.10\%}}  & {-2.67\%} \\

&{N@10}   & {0.1669}  & \multicolumn{1}{c}{0.2120}  & \multicolumn{1}{c}{0.2069}  & \multicolumn{1}{c}{0.1929}  & \multicolumn{1}{c}{0.2227}  & \multicolumn{1}{c}{0.2322}   & \multicolumn{1}{c}{\underline{0.2358}}  &{\textbf{0.2413}}  & {+2.33\%} \\ \cline{1-11}

\multirow{3}{*}{FilmTrust}   & P@10    & 25.62\%                      & 25.24\%                      & 24.37\%                      & 25.72\%                      & 0.21\%                                                  & \underline{26.19\%}        & 26.00\%                 & \textbf{27.33\%}                      & +4.35\%                     \\

& R@10               & 48.49\%                      & 46.03\%                      & 43.45\%                      &48.43\%              & 0.85\%         & \underline{49.59\%}                          & 49.11\%                                    & \textbf{51.34\%}                      & +3.53\%                     \\

& N@10                       & 0.5687                      & 0.5208                      & 0.4778                      & 0.5618                & 0.0052        & \underline{0.5805}                      & 0.5797                      & \textbf{0.5867}                      & +1.07\%                     \\ \cline{1-11} 
                        
\multirow{3}{*}{Yelp}                    & P@10                       & 1.20\%                      & 1.72\%                      & 2.19\%                      & 2.00\%                      & 2.67\%                      & 2.41\%                   & \underline{2.87\%}                      & \textbf{2.90\%}                       & +1.05\%                     \\

& R@10                       & 3.41\%                      & 4.19\%                      & 5.57\%                      & 5.27\%                      & 7.05\%                      & 6.36\%              & \underline{7.46\%}                      & \textbf{7.57\%}                      & +1.07\%                     \\
\multicolumn{1}{l}{}    & N@10                       & 0.0246                      & 0.0339                      & 0.0433                      & 0.0403                      & 0.0566                      & 0.0496                       & \underline{0.0604}                      & \textbf{0.0609}                      & +0.83\%                    \\
\bottomrule[0.3mm]
\end{tabular}
\end{table}
 

\begin{table}[]
\vspace{-1em}
\caption{ Cold-start recommendation performance result.}
\label{tab:result2}
\vspace{-1em}
\setlength\tabcolsep{2pt}

\begin{tabular}{cc|cccccccc|c}

\toprule[0.4mm]
Dataset    &{Metric} & \multicolumn{1}{c}{BPR}     & \multicolumn{1}{c}{SBPR}    & \multicolumn{1}{c}{NGCF}     & \multicolumn{1}{c}{DiffNet}    & \multicolumn{1}{c}{SGL}   & \footnotesize{$S^{2}$-MHCN}& \footnotesize{LightGCN}  & {\textbf{Motifs-Res}}     &{Improv.} \\
\midrule[0.3mm]

\multirow{3}{*}{Douban} &{P@10}   &{0.83\%} & \multicolumn{1}{c}{1.46\%} & \multicolumn{1}{c}{1.30\%} & \multicolumn{1}{c}{1.33\%} & \multicolumn{1}{c}{1.69\%} & \multicolumn{1}{c}{1.60\%}  & \multicolumn{1}{c} {\underline{1.70\%}} & {\textbf{1.74\%}} & \multicolumn{1}{c}{+2.35\%} \\

& {R@10}   & \multicolumn{1}{c}{3.62\%}  & \multicolumn{1}{c}{6.56\%}  & \multicolumn{1}{c}{5.72\%}  & \multicolumn{1}{c}{5.73\%}  & \multicolumn{1}{c}{7.31\%}   &  \multicolumn{1}{c}{7.35\%}  & \multicolumn{1}{c}{\underline{7.54\%}}  & {\textbf{7.74\%}}  & \multicolumn{1}{c}{+2.65\%} \\

& {N@10}   & \multicolumn{1}{c}{0.0252}  & \multicolumn{1}{c}{0.0468}  & \multicolumn{1}{c}{0.0407}  & \multicolumn{1}{c}{0.0417}  & \multicolumn{1}{c}{0.0504}  & \multicolumn{1}{c}{0.0521}   & \multicolumn{1}{c}{\underline{0.0541}}  &{\textbf{0.0555}}  & \multicolumn{1}{c}{+2.59\%} \\ \cline{1-11}

\multirow{3}{*}{FilmTrust}   & P@10    & 10.51\%     & 9.38\%      & 8.70\%   & 10.34\%       & 0.13\%     & 10.54\%  & \underline{10.74\%}  & \textbf{11.05\%}   & +2.89\%                     \\

& R@10     & 44.42\%        & 39.35\%      & 36.79\%    &43.66\%     & 1.02\%   & 44.70\%    & \underline{45.50\%}         & \textbf{46.93\%}        & +3.14\%                     \\

& N@10                       & 0.4612                      & 0.4012                      & 0.3554                      & 0.4502                & 0.0053        & \underline{0.4733}                      & 0.4704                      & \textbf{0.4906}                      & +3.66\%                     \\ \cline{1-11} 
                        
\multirow{3}{*}{Yelp}                    & P@10                       & 0.78\%                      & 0.98\%                      & 1.27\%                      & 1.21\%                      & 1.65\%                                     & \underline{1.73\%}           & 1.48\%               & \textbf{1.75\%}                       & +1.16\%                     \\

& R@10                       & 3.63\%                      & 4.32\%                      & 5.71\%                      & 5.45\%                      & 7.47\%      & \underline{7.87\%}                  & 6.66\%                                  & \textbf{7.95\%}                      & +1.02\%                     \\
\multicolumn{1}{l}{}    & N@10                       & 0.0.16                      & 0.0279                      & 0.0364                      & 0.0349                      & 0.0506                       & \underline{0.0525} & 0.0435                                            & \textbf{0.0535}                      & +1.90\%                    \\
\bottomrule[0.3mm]
\end{tabular}
  \vspace{-1.3em}
\end{table}
 
\begin{itemize}
\item The Motifs-Res framework performs best among all methods. Motifs-Res outperforms $S^{2}$-MHCN (the most competitive baseline) by a distinct margin, which demonstrates the performance improvement effect of the cross-motif matching representation module and the hierarchical self-supervising module proposed in this paper (this is also verified in Section 5.3.2.). LightGCN is a very competitive baseline method. Its performance is comparable to that of $S^{2}$-MHCN, and it is a lighter model with lower computational complexity.

\item 	SSL methods, such as $S^{2}$-MHCN, and Motifs-Res, overall outperform other methods, especially in the cold-start scenario, which strongly proves the beneficial effect of self-supervised learning on the recommendation performance. Various self-supervised learning tasks can autonomously mine potential characteristics from the raw data or the associated information between different motifs, which can alleviate the problem of sparsity. This is particularly important in the recommendation scenario of massive data, and it also shows that self-supervised learning is a promising approach in the field of recommender systems.

\item The GNN-based methods outperform BPR and SBPR. Graph data have more advantages than traditional euclidean data in relational modeling. The high-order relation mining ability of the hypergraph further improves the recommendation performance. This is also an important reason for the rapid industrial application of GNN-based methods in recommender systems in recent years
\item Methods that incorporate extra information such as social relationships, including SBPR, DiffNet, $S^{2}$-MHCN, and Motifs-Res, achieve better results. Social networks have become the most effective carriers for mining user interests and potential behavior characteristics. The historical behavior of users on social networks can provide the most effective reference for future recommendations. This can also explain why the performance of SGL does not meet expectations even though SGL uses self-supervised learning to address the recommendation task

\end{itemize}
 
 \subsection{Ablation Study (RQ2)}

  \begin{figure}[t]
  \centering
  \includegraphics[width=.7\linewidth,trim=0 30 0 20]{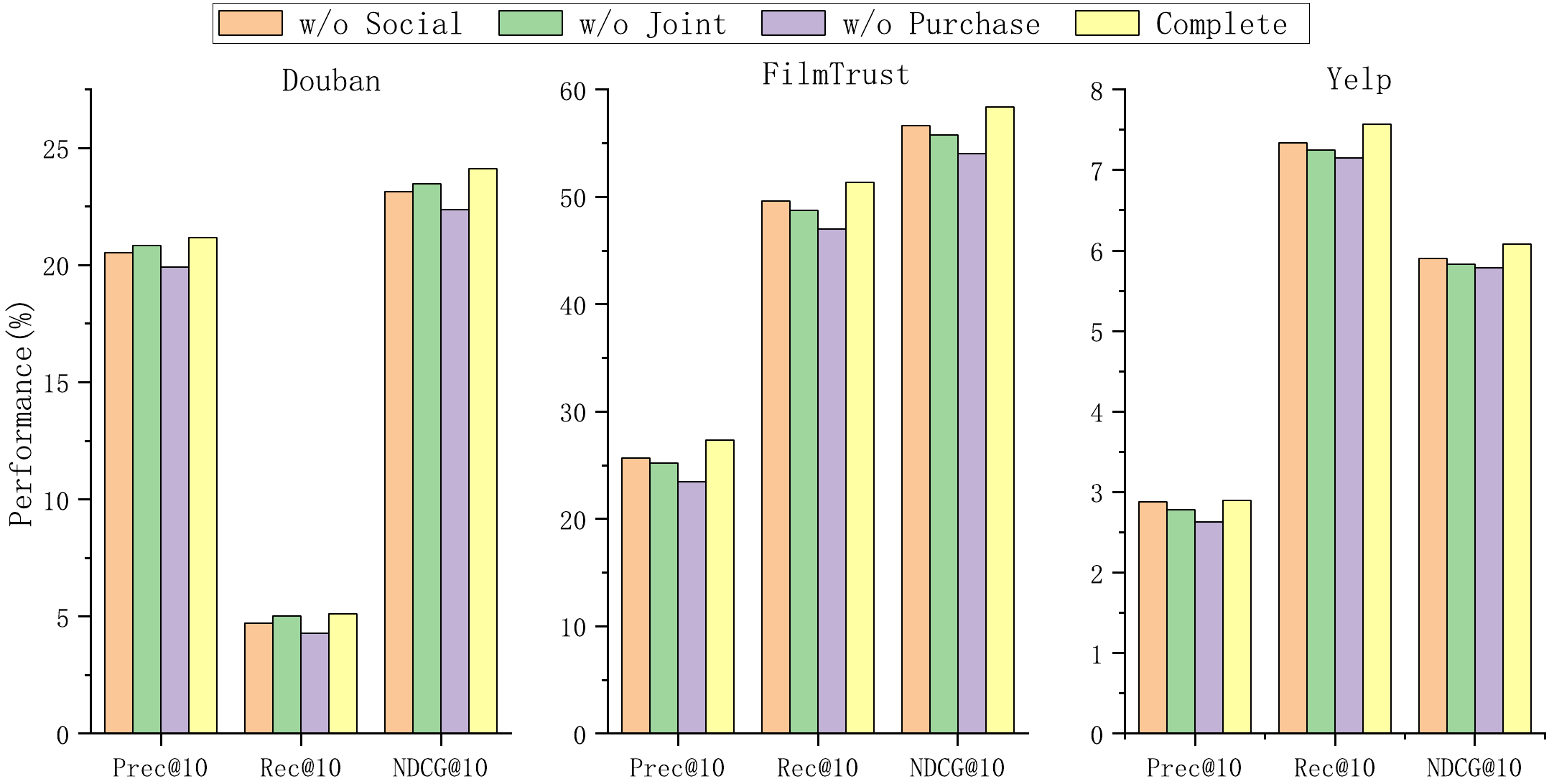}
  \caption{Plot of the contributions of information from different motifs on three datasets to the performance.}
  \vspace{-1.5em}
\end{figure}
 \subsubsection{Effects of the Multi-motif Information.} 
First, one of the highlights of this paper is the full utilization of multi-motif information for self-supervised learning to maintain the unique characteristics of different motifs and mine the associated information between different motifs. Therefore, we first conduct an ablation study on multi-motif information. Specifically, we obtain several variants by removing different motifs. For example, w/o Social denotes that the social motif has been removed and that only the joint and purchase motifs remain. By comparing the performances of different variants and the complete model, which includes all motifs, on various datasets, the contribution of each motif to the performance is evaluated. The experimental results are shown in Figure 6. By analyzing the results in Figure 6, we draw the following conclusions:
 
\begin{itemize}

\item On all datasets, the complete model consistently outperforms the three variants in which data from a single motif have been removed, which proves that the use of multiple motifs is reasonable and efficient. This conclusion is consistent with MHCN\cite{yu2021self}.

\item By comparing the three variants, we find that their performances on different datasets are different. For example, on the Douban dataset, w/o Joint has the best performance, w/o Social has the second-best performance, and w/o Purchase has the worst performance. This shows that the purchase-motif information is the most valuable on the Douban dataset, followed by the social-motif and joint-motif information. The results on the other datasets follow different trends, which shows that the information of different motifs plays different roles on different datasets. This will be further analyzed in the experiment on the attention mechanism (Figure 7).

\end{itemize}

To further evaluate the contributions of different motifs on different datasets to the performance, the attention coefficients of the motifs are recorded during the experiment. We examine the data distribution, as shown in Figure 7. Figure 7 shows that on different datasets, the information of different motifs plays different roles. On all three datasets, purchase-motif information plays the most important role. On Ciao and Yelp, most attention values of the social motif are close to 0, which is consistent with the conclusion of MHCN\cite{yu2021self}. The major difference is that the contribution of the social motif on the Douban dataset is higher than that of the joint motif. We believe that the reason may be that the addition of cross-motif matching representation and hierarchical SSL makes the model mine more important social information. By jointly analyzing Figure 6 and Figure 7, we find that the attention distributions on the datasets in Figure 7 are consistent with the performances of the corresponding variants in Figure 6. This again verifies the rationality of the multi-motif attention fusion model, namely, it demonstrates that this model can learn the attention that better integrates the information from different motifs and maximizes the performance.

\begin{figure}[t]
  \centering
  \includegraphics[width=.7\linewidth,trim=0 30 0 20]{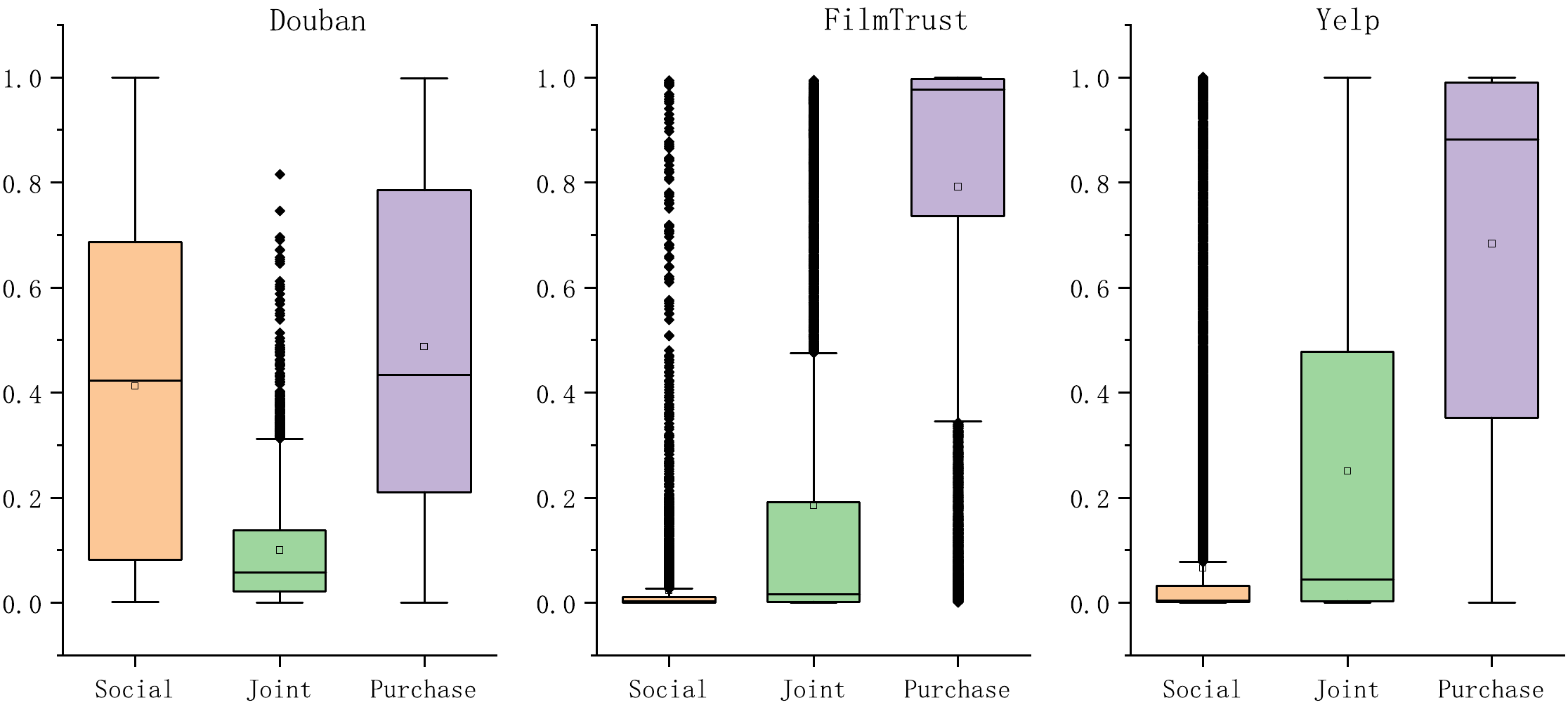}
  \caption{Attention distributions of three motifs on different datasets.}
\end{figure}

\subsubsection{Effects of the Proposed Core Components.}

To comprehensively evaluate the rationality and effects of the core components proposed in this paper, we first conduct ablation experiments and construct variants of the Motifs-Res by removing different core components. Motifs-Res$_{w/o ME\&ME.ssl}$ represents the variant that is obtained by simultaneously removing the matching representation module (Section 4.1) and the hierarchical SSL model (Section 4.2) from Motifs-Res. Motifs-Res$_{w/o ME.ssl}$ is the variant that is obtained by removing the hierarchical SSL model (Section 4.2) from Motifs-Res. In contrast to Motifs-Res$_{w/o ME\&ME.ssl}$, it contains the matching representation module (Section 4.1). Motifs-Res is the complete model, which includes both the matching representation module (Section 4.1) and the hierarchical SSL model (Section 4.2). The performances of the above three variants are shown in Figure 8. By analyzing the results in Figure 8, we obtain the following conclusions:

\begin{figure}[t]
  \centering
  \includegraphics[width=.6\linewidth,trim=0 30 0 20]{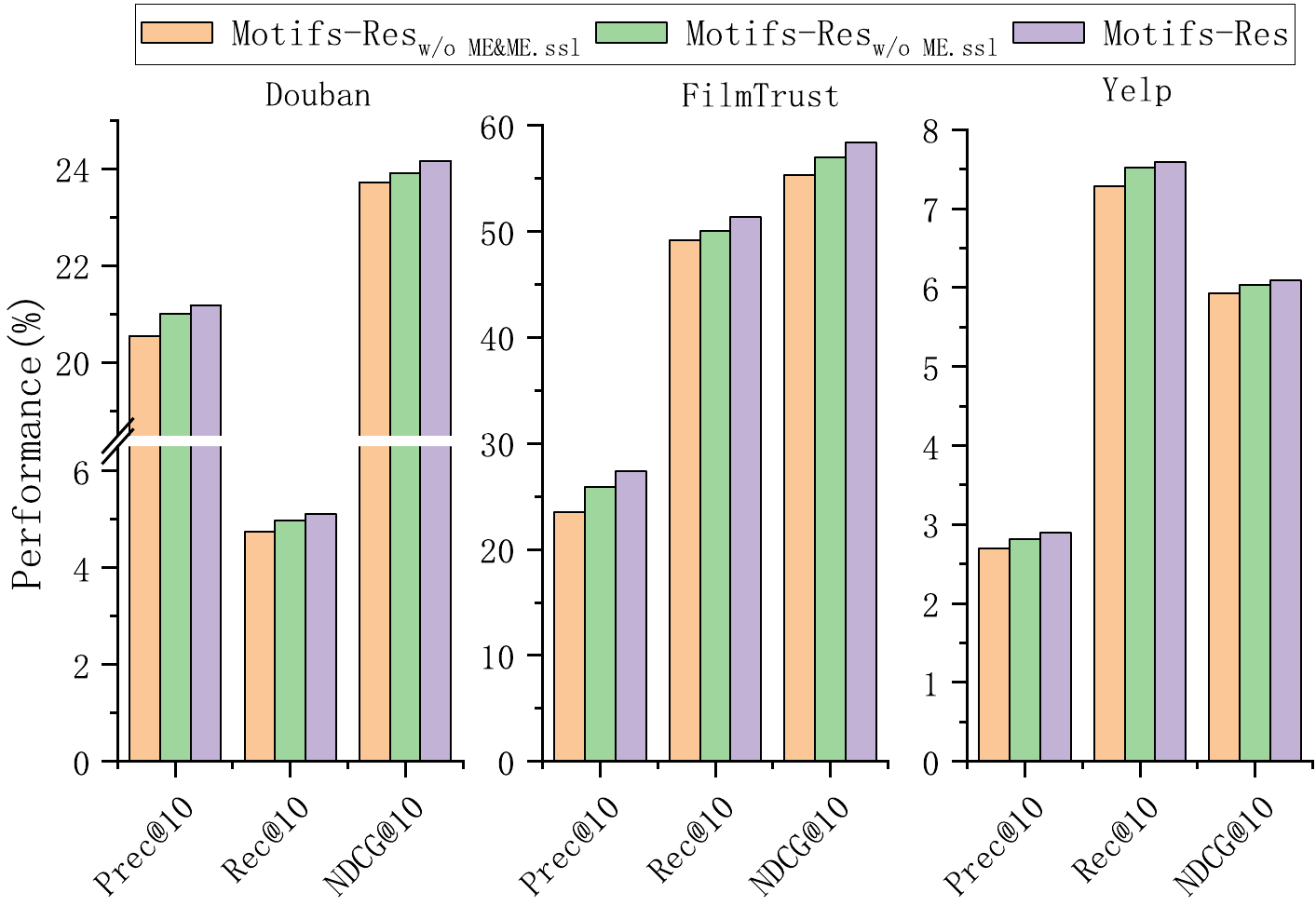}
  \caption{The performance of different variants on different datasets.}
  \vspace{-1em}
\end{figure}
\begin{itemize}

\item Compared with Motifs-Res$_{w/o ME\&ME.ssl}$, Motifs-Res$_{w/o ME.ssl}$ shows consistent and significant performance improvements on all metrics and all datasets. This shows that the matching representation module (Section 4.1) can obtain the associated information between motifs by learning the cross-motif matching representations, which is extremely valuable for recommendation tasks, even if it has not been leveraged for self-supervised learning tasks (Section 4.2).

\item 	Motifs-Res further outperforms Motifs-Res$_{w/o ME.ssl}$ on all datasets. This shows that the hierarchical self-supervised model (Section 4.2) proposed in this paper can make full use of the feature information captured by the matching representation module (Section 4.1) to establish an effective self-supervised learning task and an auxiliary task to promote the performance of the main task.

\item Through the joint analysis of Figure 8 and Figure 2, we find that a reasonable design of the matching representation model between different motifs and the self-supervision learning task based on the matching representation can not only eliminate the performance decline problem caused by directly performing contrastive learning on different motifs, but also bring significant performance improvements. This shows that the original objective of this paper has been achieved, which again verifies the exploration and analysis conclusions in Preliminaries 3.3 of this paper.

\end{itemize}

\subsection{Parameter Sensitivity Analysis (RQ3)} 

\subsubsection{Coefficients of SSL} 
In this section, we first explore the two most important hyperparameters: ${\beta _1}$, which is the coefficient of hierarchical self-supervised loss based on cross-motif matching representations (auxiliary task 1), and ${\beta _2}$, which is the coefficient of self-supervised loss based on intra-motif common representations (auxiliary task 2). Figure 9 presents the performances of different combinations of ${\beta _1}$ and ${\beta _2}$ on three datasets. On the three datasets, the overall performance of the model is relatively stable, and it has a good tolerance for the selection of ${\beta _1}$ and ${\beta _2}$. The adjustment of the two parameters does not cause significant variations in performance. Notably, Precision@10 and Recall@10 have the same performance trends as NDCG@10, due to space limitations, we do not present specific data.

\begin{figure}[h]
  \centering
  \includegraphics[width=.85\linewidth,trim=-10 30 -10 20]{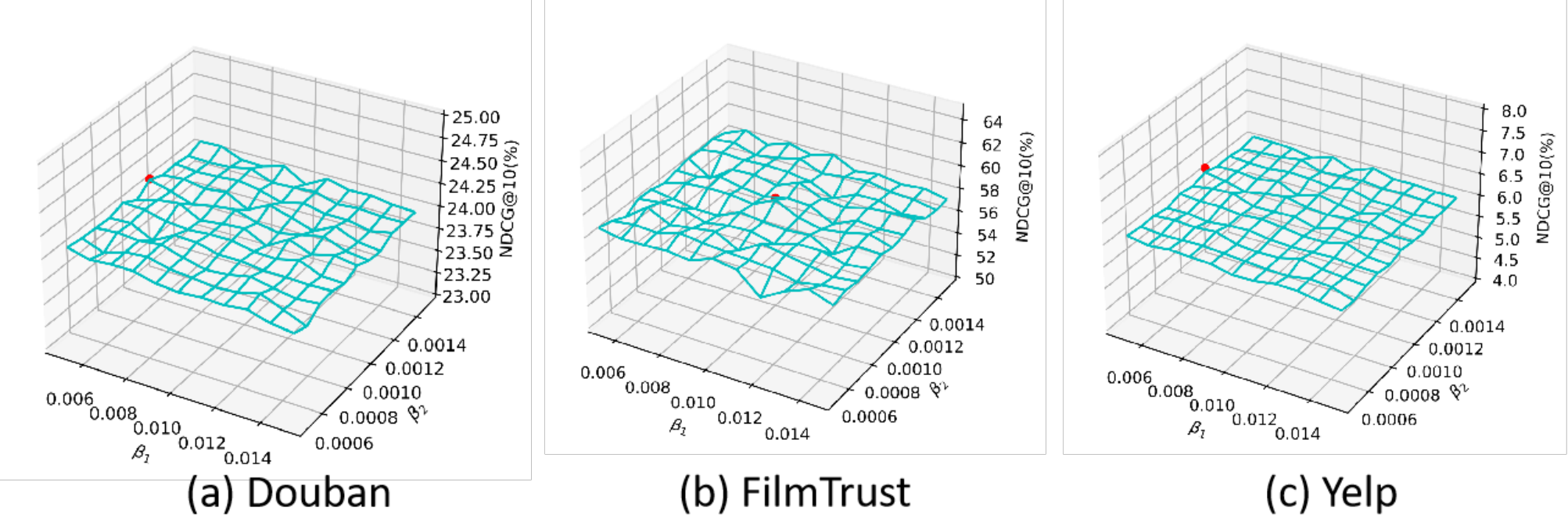}
  \caption{Performance of Motifs-Res under various combinations of ${\beta _1}$ and ${\beta _2}$ on three datasets.}
\end{figure}

\subsubsection{The depth of Motifs-Res} 

In this section, we investigate the impact of the depth of Motifs-Res. Specifically, we increase the depth from 1 to 5 and draw the performance curves of each metric on three datasets, as shown in Figure 10. The model performs best when the depth is set to 2. As the model deepens, the performance on the Ciao dataset declines more obviously, whereas the performances on the Douban and Yelp datasets are relatively stable. However, the performances on the three datasets consistently decline, which is consistent with the findings of MHCN\cite{yu2021self}. The reason is that our proposed framework, which is based on motifs, naturally extracts the high-order neighborhoods of nodes. Compared with the ordinary GNN model, as the model deepens, the overfitting phenomenon occurs earlier. This is also in line with the conclusions regarding the depth of the model in previous studies\cite{sun2022mhnf,li2021deepgcns}. Overfitting caused by deep models is also a problem that needs to be further studied in future works on GNNs, especially Hyper-GNNs.

\begin{figure}[h]
  \centering
  \includegraphics[width=.7\linewidth,trim=0 30 0 20]{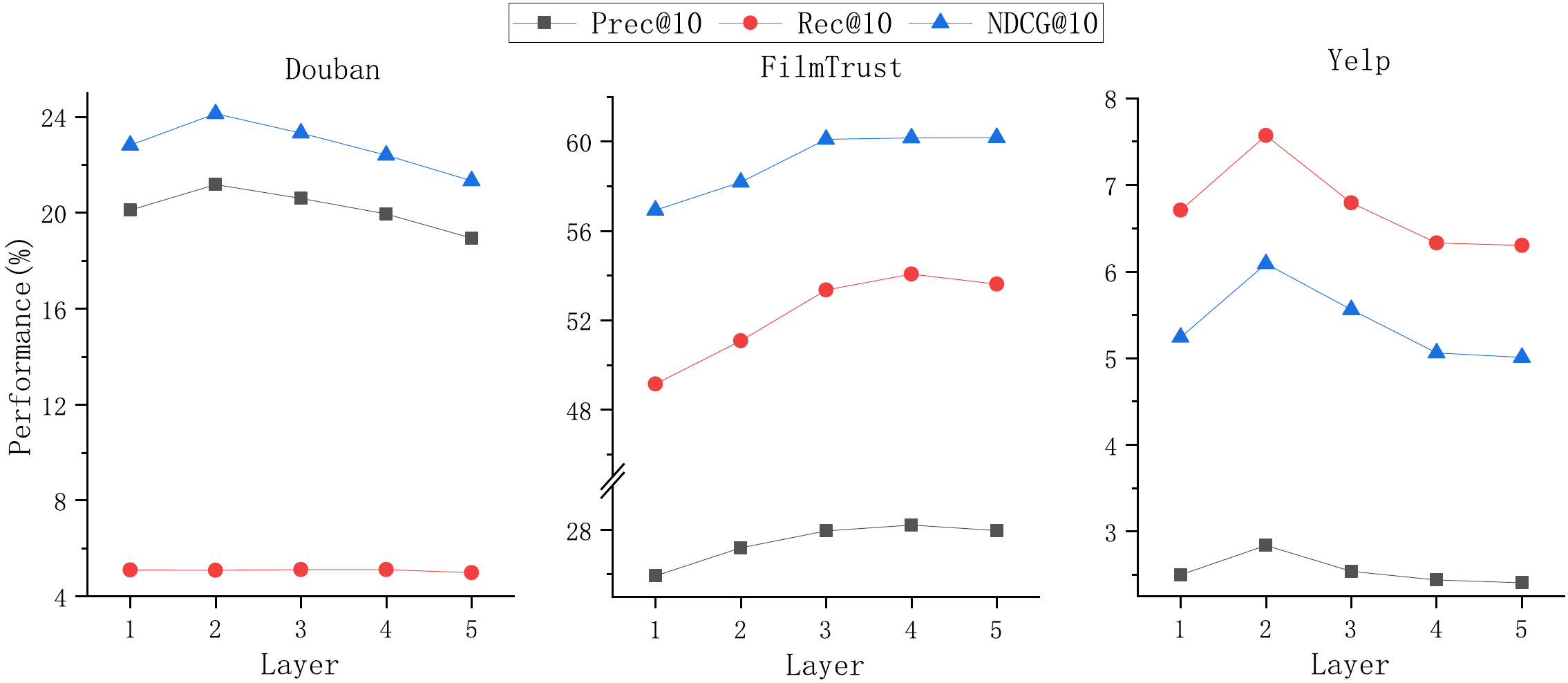}
  \caption{Performance of Motifs-Res with various depths on three datasets.}
\end{figure}

\subsubsection{Temperature Hyperparameter of InfoNCE} 

Finally, we also investigate the impact of the $\tau $ value on the performance of Motifs-Res. Figure 11 shows that when $\tau $ is set too low (below 0.05), all metrics on each dataset drop dramatically. When $\tau $ is in the range of [0.05, 0.8], the performance of the model is better and more stable, which indicates that this range is reasonable. This conclusion is consistent with the findings of SGL\cite{wu2021self}.

\begin{figure}[h]
  \centering
  \includegraphics[width=.7\linewidth,trim=0 30 0 20]{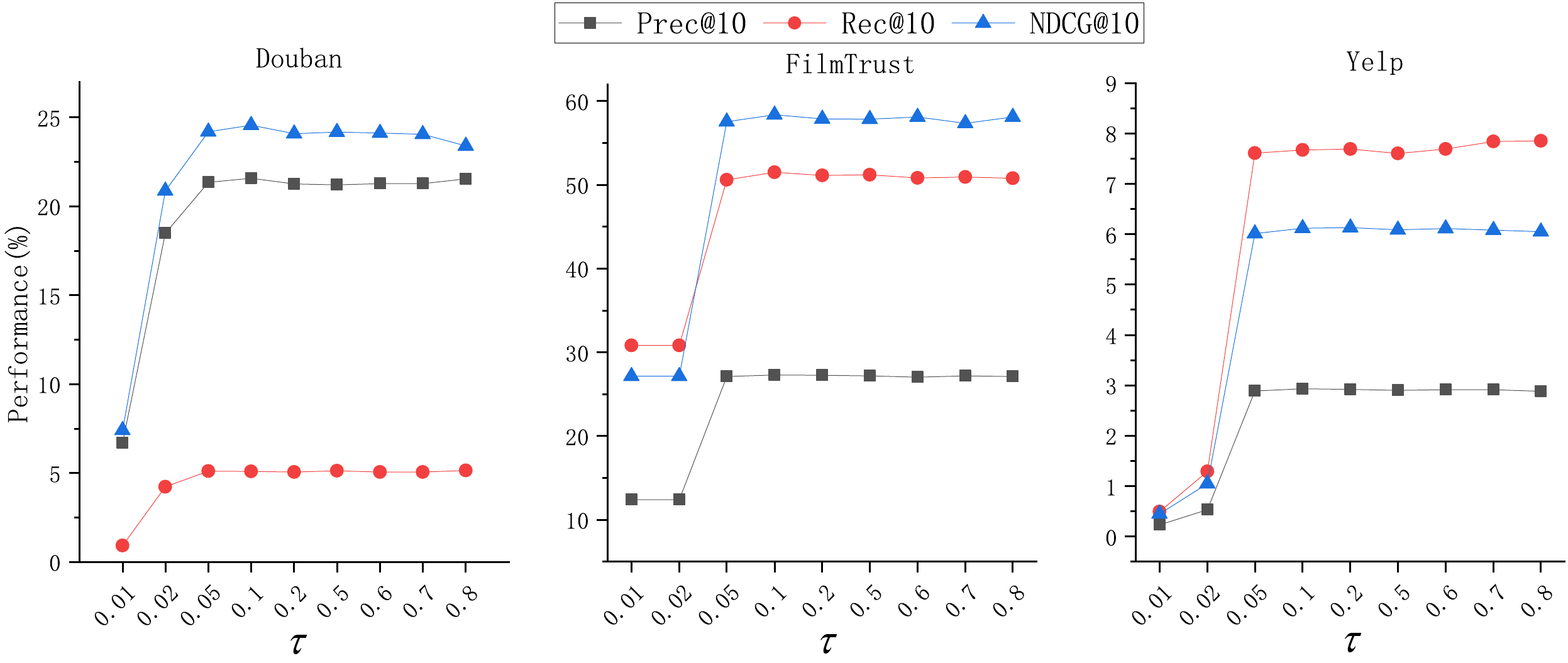}
  \caption{Performance of Motifs-Res with various $\tau $ values on three datasets.}
\end{figure}

\section{Conclusion and Future Work} 

The key issue studied in this paper is leveraging multi-motif data to perform more efficient self-supervised learning tasks and then enhance the recommendation performance. First, we assumed that directly constructing a contrastive learning task on the features of different motifs would make the data of each motif homogeneous, which would deteriorate the recommendation performance. In the exploration experiment, we verified this assumption. To tackle this problem, we proposed the Motifs-Res framework, which can make full use of the information in each motif and the information across motifs to construct self-supervised learning tasks, thereby improving recommendation performance. Specifically, to comprehensively mine the associated information between motifs while avoiding the problem of homogenization, we proposed a cross-motif matching representation learning model, in which a cross-motif matching representation is learned by attentive matching. On this basis, we innovatively proposed a cross-motif hierarchical SSL model based on matching representation, which realizes self-supervised learning within and between motifs, respectively, and improves the ability of self-supervised learning tasks to autonomously mine different levels of potential information. Finally, we unified the recommendation task (main task), hierarchical self-supervised learning task(auxiliary task 1), and self-supervised learning task based on intra-motif common representations (auxiliary task 2) for joint learning. Extensive experimental results on three real datasets showed that Motifs-Res outperforms the state-of-the-art methods by large margins, and the ablation studies also proved the benefits of each core component proposed in this paper.

However, in the model depth exploration experiment (Section 5.4.2), we also found that as the model deepens, the performance on some datasets drops significantly. The reason was identified in Section 5.4.2: our proposed framework, which is based on motifs, naturally extracts the high-order neighborhood of nodes. Compared with the ordinary GNN model, as the model deepens, the overfitting phenomenon occurs earlier. This is also the main disadvantage of the hypergraph neural network that needs to be addressed \cite{sun2022mhnf,li2021deepgcns}. Therefore, overfitting caused by the deep model is also a problem that needs to be further studied in future works on graph neural networks, especially hypergraph neural networks. In the meantime, integrating valuable information, such as text, sound, and images, is also a promising research direction in recommender systems.

\begin{acks}
This work is supported by the Fundamental Research Funds for the Central Universities (grant no. HIT.NSRIF.201714), Weihai Science and Technology Development Program (2016DXGJMS15), and the Key Research and Development Program in Shandong Province (2017GGX90103).
\end{acks}

\bibliographystyle{ACM-Reference-Format}
\bibliography{./bib}
\end{document}